\documentclass[journal]{IEEEtran}
\usepackage[english]{babel}
\usepackage{blindtext}
\usepackage{pstricks}
\usepackage{amsmath,amssymb,amscd,latexsym}
\usepackage{breqn}
\usepackage{dashrule}
\usepackage{comment,psfrag,subfigure,enumerate}
\usepackage{cancel}
\usepackage{cite}
\usepackage{url}
\usepackage{tensor}
\usepackage{graphicx}
\usepackage{pstool}
\usepackage{float,color}
\usepackage{rotating}
\usepackage{multirow}
\usepackage{multicol}
\usepackage{setspace}
\usepackage{hyperref}
\usepackage{graphicx} 
\usepackage{subfigure}
\usepackage{times}
\usepackage{tikz}
\usepackage{verbatim}
\usepackage{pst-optexp}
\usepackage[framemethod=TikZ]{mdframed}
\usepackage[utf8]{inputenc}
\usepackage[T1]{fontenc}
\usepackage{tabularx, makecell, booktabs}%
\usepackage{etoolbox}
\setcellgapes{3pt}
\usepackage{pgfplotstable,enumerate}
\usetikzlibrary[pgfplots.groupplots]
\usepgflibrary{decorations.pathmorphing} 
\usetikzlibrary{arrows.meta}
\usetikzlibrary{arrows,shapes}
\usetikzlibrary{arrows,patterns}
\usetikzlibrary{spy}

\newcommand{\Rmnum}[1]{\expandafter\@slowromancap\romannumeral #1@}

\usetikzlibrary{calc}
\mdfdefinestyle{MyFrame}{%
    usetwoside=false,
    linecolor=black,
    outerlinewidth=0.5pt,
    roundcorner=10pt,
    innertopmargin=\baselineskip,
    innerbottommargin=\baselineskip,
    innerrightmargin=10pt,
    leftmargin=10pt,
    rightmargin=10pt,
    skipabove = -10pt,
    skipbelow = 0pt,
    innerleftmargin=10pt,
    userdefinedwidth = 1.0\textwidth
    backgroundcolor=gray!1!white}
    
    \mdfdefinestyle{MyFrame1}{%
    usetwoside=false,
    linecolor=black,
    outerlinewidth=0.5pt,
    roundcorner=10pt,
    innertopmargin=\baselineskip,
    innerbottommargin=\baselineskip,
    innerrightmargin=10pt,
    leftmargin=10pt,
    rightmargin=10pt,
    skipabove = 10pt,
    skipbelow = 0pt,
    innerleftmargin=10pt,
    userdefinedwidth = 1.0\columnwidth
    backgroundcolor=gray!1!white}

\newcolumntype{f}{>{$}l<{$}}
\newcolumntype{n}{l}
\newcolumntype{N}{>{\scriptsize}l}
\newcolumntype{v}[1]{>{\raggedright\hspace{0pt}}p{#1}}
\newcolumntype{V}[1]{>{\scriptsize\raggedright\hspace{0pt}}p{#1}}
\newcolumntype{B}[1]{>{\boldmath\DC@{.}{,}{#1}}l<{\DC@end}}
\newcolumntype{d}[1]{>{\DC@{.}{,}{#1}}l<{\DC@end}}
\newcolumntype{i}[1]{>{\DC@{.}{,}{#1}\mathnormal\bgroup}l<{\egroup\DC@end}}
\newcolumntype{s}[1]{>{\DC@{.}{,}{#1}\mathsf\bgroup}l<{\egroup\DC@end}}
\newcolumntype{R}[1]{%
  >{\begin{turn}{90}\begin{minipage}{#1}\scriptsize\raggedright\hspace{0pt}}l%
  <{\end{minipage}\end{turn}}%
}
\newcolumntype{x}{>{\scriptsize\raggedright\hspace{0pt}}X}
\makeatother

\usetikzlibrary{intersections,arrows,decorations.pathmorphing,snakes}

\pgfdeclaredecoration{fiber}{initial}
{
    \state{initial}[width=1.5\pgfdecorationsegmentamplitude, next state=final]{
    \pgfpathmoveto{\pgfpointorigin}
    \pgflineto{\pgfpoint{\pgfdecoratedinputsegmentlength/2}{0pt}}
    \pgfpathcircle{\pgfpoint{\pgfdecoratedinputsegmentlength/2}{.5\pgfdecorationsegmentamplitude}}{.5\pgfdecorationsegmentamplitude}
    \pgfpathcircle{\pgfpoint{\pgfdecoratedinputsegmentlength/2-.25\pgfdecorationsegmentamplitude}{.5\pgfdecorationsegmentamplitude}}{.5\pgfdecorationsegmentamplitude}
    \pgfpathcircle{\pgfpoint{\pgfdecoratedinputsegmentlength/2+.25\pgfdecorationsegmentamplitude}{.5\pgfdecorationsegmentamplitude}}{.5\pgfdecorationsegmentamplitude}
    \pgfpathmoveto{\pgfpoint{\pgfdecoratedinputsegmentlength/2}{0pt}}
    }
   \state{final}{
        \pgfpathlineto{\pgfpointdecoratedpathlast}
   }
}

\begin{document}

\title{Analytical Modeling of Nonlinear Fiber Propagation for Four Dimensional Symmetric Constellations}
\author{Hami~Rabbani, Mostafa Ayaz, Lotfollah~Beygi, Gabriele Liga, {\it{Member}}, {\it{IEEE}}, Alex Alvarado, {\it{Senior Member}}, {\it{IEEE}}, Erik Agrell, {\it{Fellow}}, {\it{IEEE}},  Magnus Karlsson, {\it{Senior Member}}, {\it{IEEE}}, {\it{Fellow}},  {\it{OSA}}
\thanks{H.~Rabbani, M.~Ayaz and L.~Beygi are with the the EE Dept. of K. N. Toosi University of Technology. E-mails: h.ami.rabbani@email.kntu.ac.ir, m.ayaz@email.kntu.ac.ir and beygi@kntu.ac.ir}
\thanks{G.~Liga and A.~Alvarado are with the Information and Communication Theory Lab, Signal Processing Systems Group, Department of Electrical Engineering, Eindhoven University of Technology, Eindhoven 5600 MB, The Netherlands. E-mails:
\{g.liga,a.alvarado\}@tue.nl
}
\thanks{E.~Agrell is with the Dept. of Electrical Engineering, Chalmers University of Technology, Sweden. E-mail: agrell@chalmers.se}
\thanks{M.~Karlsson is with the Dept. of Microtechnology and Nanoscience, Photonics Laboratory, Chalmers University of Technology, Sweden. E-mail: magnus.karlsson@chalmers.se}
\thanks{G.~Liga is funded by the EuroTechPostdoc programme under the European Union’s Horizon 2020 research and innovation programme (Marie Skłodowska-Curie grant agreement No.~754462). This work has received funding from the European Research Council (ERC) under the European Union’s Horizon 2020 research and innovation programme (grant agreement No. 757791).}
}

\markboth{Preprint, \today}{}%
\maketitle

\newpage

\begin{abstract}
Coherent optical transmission systems naturally lead to a four dimensional (4D) signal space, i.e., two polarizations each with two quadratures. In this paper we derive an anlaytical model to quantify the impact of Kerr nonlinearity on such 4D spaces, taking the interpolarization dependency into account. This is in contrast to previous models such as the GN and EGN models, which are valid for polarization multiplexed (PM) formats, where the two polarizations are seen as independent channels on which data is multiplexed. The proposed model
agrees with the EGN model in the special case of independent two-dimensional modulation in each polarization.
The model accounts for the predominant nonlinear terms in a WDM system, namely self-phase modulation and and cross-phase modulation.
Numerical results show that the EGN model may inaccurately estimate the nonlinear interference of 4D formats. This nonlinear interference discrepancy between the results of the proposed model and the EGN model could be up to 2.8 dB for a system with 80 WDM channels. The derived model is validated by split-step Fourier simulations, and it is shown to follow simulations very closely.
\end{abstract}



\section{Introduction}

\IEEEPARstart{T}{he} amount of traffic carried on optical backbone networks continues to grow at a rapid pace, and makes efficient use of available resources indispensable. The Kerr nonlinearity is the overriding factor that leads to signal distortion and 
limits the capacity of optical fiber transmission systems \cite{Essiambre_2010}. Studying the ultimate limits of such systems is key to avoid the capacity crunch. To circumvent the capacity cruch problem, spectrally-efficient modulation formats have attracted substantial attention.  

Optimized 2D modulation formats have become increasingly popular in optical communications. 
However, further optimization is possible if the full 4D signal space (which is inherent in optical coherent detection) is exploited.
The idea of 4D modulation formats was introduced to optical communications as far back in time as the coherent receiver was explored \cite{betti1990exploiting, betti1991novel,benedetto1992theory,cusani1992efficient}. Agrell and Karlsson \cite{Erik.optimized.modulation,karlsson2009most} began optimizing modulation formats in a 4D space for coherent optical communication systems in 2009. A number of 4D modulation formats have recently been proposed for purposes of maximizing generalized mutual information, optimizing power efficiency, and other equally compelling motivations \cite{Kojima2017,Reimer,Nakamura, Chen2019}. 4D coded modulation with bit-wise decoders was studied in \cite{Alvarado2015}. Recently, other 4D coded modulation schemes have been proposed in \cite{Cai2020,Frey2020}.

Although quite a few approximate analytical models for nonlinear fibre propagation are currently available in the literature\cite{ CarenaGaussian2012, A.Mecozzi2012, Pontus_JLT_modeling_2013, dar2013properties, Curri2013, carena2014egn}, all of these models aim to predict the nonlinear interference (NLI) in polarization multiplexed (PM) systems. 
What follows is a short description of analytical models proposed for such PM optical systems.


To analytically evaluate the quality of transmissions of fiber-optic links, many research works have been devoted to extracting channel models both in the time and frequency domains  \cite{Carena2012, A.Mecozzi2012,
carena2014egn}.
The Gaussian noise (GN) models in highly dispersive optical communications systems were presented in \cite{Pontus_JLT_modeling_2013, Carena2012,Poggiolini2012, serena2013alternative}. The 4D GN-type channel model was first proposed in \cite{Beygi_2012}. The finite-memory GN model was introduced in \cite{Erik.Finite.memory.model}.
Due to the Gaussianity assumption of the signal, GN model is not able to predict the modulation format dependence property of NLI. 

The authors of \cite{A.Mecozzi2012} for the first time addressed a modulation-format-dependent time-domain model, assuming only the
dominant nonlinear terms of cross-channel interference (XCI),
known as cross-phase modulation (XPM) terms. Later, this
time-domain model was studied comprehensively in \cite{dar2013properties} and
compared with the GN model to address the discrepancy
between these two models.
In much the same way as in \cite{dar2013properties}, the authors of \cite{carena2014egn} derived a new perturbation model (in the frequency domain) dropping the assumption of Gaussianity of the transmitted signal. This model was labelled enhanced Gaussian noise (EGN) model. As its name suggests, the EGN model added a number of correction terms to the GN model formulation, which fully captured the modulation format dependency of the NLI. Moreover, the frequency-domain approach in \cite{carena2014egn} allows the model to fully account for all the different contributions of the NLI in a WDM spectrum, including: the self-channel interference (SCI), and unlike \cite{dar2013properties}, all XCI and multi-channel interference (MCI) terms. 
It was shown in \cite{poggiolini2016analytical} that the GN and time-domain model \cite{A.Mecozzi2012,dar2013properties} failed to accurately predict the NLI, whilst the EGN model was able to capture both the modulation format and the symbol rate dependency of the NLI. 
The achievable rate in nonlinear WDM systems was evaluated in \cite{secondini2013achievable}.

Recently, \cite{Semrau.2019.MD.jlt} proposed a modulation-format-dependent model in the presence of stimulated Raman scattering. The authors of \cite{Semrau.2019.MD.jlt} added a modulation format correction term to XPM, while SCI was computed under a
Gaussian assumption. A general nonlinear model in the presence of Kerr nonlinearity and stimulated Raman scattering was proposed in \cite{rabbani2019general}, which accounts for the modulation-format-dependent SCI, XCI, and MCI terms. A survey of channel models proposed in the literature up to 2015 was presented in \cite{Erik.Survey}.

All of the aforementioned works are valid for PM modulation formats in which polarizations act as two independent channels.
In this paper, we concentrate on symmetric\footnote{Constellations which are symmetric with respect to the origin, and have the same power in both polarizations.} constellations and derive an accurate analytical model that is able to predict the impact of NLI on 4D optical transmission systems where data is jointly transmitted on both polarizations. Unlike the previous models \cite{carena2014egn, dar2013properties, golani2016modeling}, the derived model is built on the fact that the x- and y-polarization are dependent of one another, making it possible to predict the performance of optimized 4D modulation formats in the presence of fiber nonlinearities. A comprehensive approach to deriving the SCI term in the frequency domain is currently being developed in \cite{liga2020extending}, thus enabling the computation of the NLI power of arbitrary zero-mean 4D constellations.

The paper computes the SCI and XPM nonlinear
terms.
Our model is derived following a time-domain approach, as in \cite{MecozziEssiambre2012, dar2013properties}, and does not include other XCI terms apart from XPM, nor does it contain MCI \cite[Fig.~7]{carena2014egn}.
Although the derivation of a comprehensive analytical model that can take into account all terms of NLI (SCI, XCI, and MCI) goes beyond the scope of this paper, the model in this paper computes the lion's share of the NLI in multi-channel WDM systems, i.e., the SCI and XPM terms \cite[Fig. 2]{poggiolini2016analytical}. 

The rest of this paper is organized as follows. In Sec. \ref{preliminaries}, we
 describe the electrical field in a 4D space and also review the first order solution to Manakov equation. The main result of this work is presented in Sec.~\ref{key.results}. 
In Sec.~\ref{numericalSec}, we validate the proposed model by split-step Fourier simulations, and compare a wide variety of 4D formats in terms of the experienced NLI. Sec.~\ref{conclusion} concludes the paper. The detailed derivations of the main result of this paper are included in the Appendix.


\section{preliminaries}\label{preliminaries}

The electric field of the optical wave intrinsically comprises two polarizations, each with two quadratures, thus in total four degrees of freedom, any one of which can be considered as a dimension. The electrical field can therefore be written as{\footnote{Throughout this paper we use $(\cdot)_{\text{x}}$ and $(\cdot)_{\text{y}}$ to represent variables associated to polarizations $\text{x}$ and $\text{y}$, resp.
Expectations are denoted by $\mathbb{E}\{\cdot\}$, and two dimensional complex functions are denoted using boldface (e.g., $\boldsymbol{E}$) symbols whose Hermitian conjugate is shown by $(\cdot)^{\dagger}$.}}
\begin{align}\label{elec.field}
     \boldsymbol{E}&= \begin{bmatrix}
           E_{\text{x}} \\
           E_{\text{y}} \\
         \end{bmatrix}=\begin{bmatrix}
           E_{\text{x},\text{r}}+iE_{\text{x},\text{i}} \\
           E_{\text{y},\text{r}}+iE_{\text{y},\text{i}} \\
         \end{bmatrix},
\end{align}
where indices x and y stand for polarization states, and r and i the real and imaginary parts, resp., of the electrical field.

The propagation of dual-polarized signals in a dispersive and nonlinear optical fiber is governed
by the Manakov equation \cite[Ch.~2]{agrawal_linear_2006}
\begin{align}
    \label{Manakov}
    \frac{\partial}{\partial z}\!\boldsymbol{E}(t,z)\!=\!\!-\frac{i\beta_2}{2}\frac{\partial^2}{\partial t ^2}\boldsymbol{E}(t,z)\!+\!i\frac{8}{9}\gamma f(z)\boldsymbol{E}^\dagger(t,\!z)\boldsymbol{E}(t,\!z)\boldsymbol{E}(t,\!z),
\end{align}
where $\gamma$ is the nonlinearity
coefficient, $\beta_2$ is the group velocity dispersion, and  $f(z)$ accounts for the
link’s loss/gain profile. In
the case of perfectly distributed amplification $f(z)=1$, while in the case of lumped amplification $f(z)=\text{exp}\{-\alpha\text{mod}(z,L)\}$ where $\alpha $ is the loss coefficient, $L$ is the span length and
$\text{mod}(z,L)$ is the modulo operation and shows the distance between the point $z$ and the nearest preceding amplifier.

We wish to evaluate the variance of SCI (intra-channel interference) and XPM (inter-channel interference) terms based on the first order perturbation approach, as these terms contribute to the NLI as predominant factors. We consider a channel of interest (COI) whose central frequency is set to zero, and an interfering channel with central frequency $\Omega$. The XPM contributions of multiple WDM channels sum up incoherently, so there is no need to consider more than one channel pair \cite[Sec. 2]{dar2013properties}. The linear solution of the Manakov equation at distance $z$ for two channels is \cite[Eq.~(1)]{dar2013properties}
\begin{align}\label{zeroth.order solution}
 \boldsymbol{E}(z,t)=&\sum_{k}\boldsymbol{a}_kg(t-kT,z)\nonumber\\&+ \text{e}^{-i\Omega t+\frac{i\beta_2\Omega^2}{2}z}\sum_{k}\boldsymbol{b}_k g(t-kT-\beta_2\Omega z,z),
\end{align}
where $\boldsymbol{a}_k=[a_{k,\text{x}}\;a_{k,\text{x}}]^{\text{T}}$ and $\boldsymbol{b}_k=[b_{k,\text{x}}\; b_{k,\text{x}}]^{\text{T}}$ are column vectors containing two elements, which represent the $k$-th symbol transmitted by the COI and interfering channel, resp. The dispersed pulse is represented by $g(t,z)=\text{exp}(-iz\beta_2\partial_t^2/2)g(t,0)$ \cite{dar2015inter}, where $g(t,0)$ is the input pulse, and $\partial_t^2$ is the time derivative operator. The symbol rate of channels is denoted by $T^{-1}$. 

Without loss of generality, we concentrate on detecting the zeroth symbol in the COI, i.e., $\boldsymbol{a}_0$. The receiver for the COI is assumed to fully compensate for the linear link's impairments. The received symbol at the end of the link is therefore expressed as $\boldsymbol{a}_0+\Delta \boldsymbol{a}_0$, where $\Delta \boldsymbol{a}_0$ is the NLI contribution.
The first order solution to Manakov equation is obtained based on the perturbation approach \cite[Eq.~(3)]{dar2015inter}, which gives
\begin{align}
    \label{first.order.solution}
    \Delta\boldsymbol{a}_0(\Omega)=&i\frac{8}{9}\gamma\sum_{h,k,l}S_{h,k,l}\boldsymbol{a}^{\dagger}_{k}\boldsymbol{a}_{h}\boldsymbol{a}_{l}\nonumber\\&+i\frac{8}{9}\gamma\sum_{h,k,l}\!\!X_{h,k,l}\left(\boldsymbol{b}^{\dagger}_{k}\boldsymbol{b}_{h}\mathbb{I}+\boldsymbol{b}_{h}\boldsymbol{b}^{\dagger}_{k}\right)\boldsymbol{a}_{l}.
\end{align}
In \eqref{first.order.solution} $\mathbb{I}$ is the $2\times2$ identity matrix, and $S_{h,k,l}$ and $X_{h,k,l}$ are  \cite[Eqs.~(4) and (5)]{dar2015inter}
\begin{align}\label{coeficient.S}
S_{h,k,l}=&\int_{0}^{L}\text{d}z\int_{-\infty}^{\infty}\text{d}t f(z)g^*(t,z)g(t-lT,z)\nonumber\\&\cdot g^*(t-kT,z)g(t-hT,z),
\end{align}
and
\begin{align}\label{coeficient.X} 
X_{h,k,l}=&\int_{0}^{L}\text{d}z\int_{-\infty}^{\infty}\text{d}t f(z)g^*(t,z)g(t-lT,z)\nonumber\\&\cdot g^*(t-kT-\beta_2\Omega z,z)g(t-hT-\beta_2\Omega z,z),
\end{align}
resp. The first and second terms on the right-hand side of \eqref{first.order.solution} are responsible for estimating the SCI and XPM terms, resp. Using the fact that $g(t,z)=\int\text{d}w\tilde{g}(w)\text{exp}(-iwt+iw^2\beta_2z/2)/(2\pi)$, where $\tilde{g}(w)$ is the Fourier transform of $g(t,0)$ (see \cite[Appendix]{dar2015inter} and \cite[Eqs.~(11) and (12)]{dar2015inter}), \eqref{coeficient.S} and \eqref{coeficient.X} are expressed in the frequency domain as
\begin{align}
    \label{S.coef.alt.freq}
    S_{h,k,l}=\int\frac{\text{d}^3w}{(2\pi)^3} \rho_{\text{s}}(w_1,w_2,w_3)\text{e}^{i(w_1h-w_2k+w_3l)T},
\end{align}
and
\begin{align}
    \label{X.coef.alt.freq}
    X_{h,k,l}=\int\frac{\text{d}^3w}{(2\pi)^3} \rho_{\text{xp}}(w_1,w_2,w_3)\text{e}^{i(w_1h-w_2k+w_3l)T},
\end{align}
resp., where $\int{\text{d}^3w}$ stands for $\int_{-R/2}^{R/2}\int_{-R/2}^{R/2}\int_{-R/2}^{R/2} \text{d}w_1\text{d}w_2\text{d}w_3$ in which $R=2\pi/T$, and
\begin{align}
\label{rho.s}
&\rho_{\text{s}}(w_1,w_2,w_3)=\tilde{g}^*(w_1-w_2+w_3)\nonumber\\&\cdot\tilde{g}(w_1)\tilde{g}^*(w_2)\tilde{g}(w_3)\int_{0}^{L}\text{d}zf(z)\text{e}^{i\beta_2(w_2-w_3)(w_2-w_1)z},
\end{align}
and
\begin{align}
\label{rho}
&\rho_{\text{xp}}(w_1,w_2,w_3)=\tilde{g}^*(w_1-w_2+w_3)\nonumber\\&\cdot\tilde{g}(w_1)\tilde{g}^*(w_2)\tilde{g}(w_3)\int_{0}^{L}\text{d}zf(z)\text{e}^{i\beta_2(w_2-w_3+\Omega)(w_2-w_1)z}.
\end{align}
One may want to take all the NLI terms such as SCI, XCI and MCI into account. In this regard, \eqref{zeroth.order solution} should be extended to a general equation, which accounts for $N$ terms, where $N$ is the number of WDM channels occupying the full C-band spectrum, and as a result, 
\eqref{first.order.solution} will contain $N^{3}$ terms, which stem from $\boldsymbol{E}^\dagger(t,z)\boldsymbol{E}(t,z)\boldsymbol{E}(t,z)$ in \eqref{Manakov}. 
Nonlinear analysis of all the NLI terms however falls outside the scope of the paper and is left for future work.
\section{The key result: NLI variance}\label{key.results}

This section is devoted to providing the key result of this work, which is the variance of \eqref{first.order.solution}. 
Not only is the key result able to predict the NLI of most 4D constellations used in practice, it is straightforward enough to be easily calculated with even the simplest of computers. The detailed derivation of the key result will be given in the Appendix. The key result is obtained under some simplifying assumptions, which are discussed below.

The first assumption is that the data symbols in the x- and y-polarization are correlated with each other. The second assumption is that the data symbols in different time slots are independent of one another.
Here, we consider a multi-channel WDM system 
where channels across the spectrum can have different launch powers and different 4D modulation formats. The probability distribution in each WDM channel is assumed uniform over all constellation points. We further assume that the launch power in the x- and y-polarization are the same, meaning that
\begin{align}\label{half.power}
&\frac{P_{\text{\tiny COI}}}{2}=\mathbb{E}\{|a_{\text{x}}|^
2\}=\mathbb{E}\{|a_{\text{y}}|^2\},\quad \frac{P_{\text{\tiny INT}}}{2}=\mathbb{E}\{|b_{\text{x}}|^2\}=\mathbb{E}\{|b_{\text{y}}|^2\},
\end{align}
where $P_{\text{\tiny COI}}$ and $P_{\text{\tiny INT}}$ are the total launch power transmitted in the COI and interfering channel, resp. It is also assumed that 
\begin{align}
    \label{fourth.order.moment}
   \mathbb{E}\{|a_{\text{x}}|^4\}=\mathbb{E}\{|a_{\text{y}}|^4\},\quad \mathbb{E}\{|b_{\text{x}}|^4\}=\mathbb{E}\{|b_{\text{y}}|^4\}.
\end{align}
The last key assumption is that $\mathbb{E}\{a_{\text{x}}\}=\mathbb{E}\{a_{\text{y}}\}=\mathbb{E}\{a_{\text{x}}^2\}=\mathbb{E}\{a_{\text{x}}a_{\text{y}}^*\}=\mathbb{E}\{|a_{\text{x}}|^2a_{\text{x}}\}=\mathbb{E}\{|a_{\text{y}}|^2a_{\text{x}}\}=0$. This assumption holds for most zero-mean symmetric constellations with respect to the origin that have the same power in both polarizations.
Although we will show the NLI variance for Nyquist rectangular spectral shape channels (sinc pulse), the results can also be used for near rectangular signal
spectral shape such as a root raised cosine with small roll off factor. 

The NLI variance on the $n$-th channel (COI) caused by \eqref{first.order.solution} is given by
\begin{align}
    \label{nli.multi.channel}
    \sigma_{\text{NLI},n}^2=&\mathrm{Var}\left\{\sum_{\Omega}\Delta\boldsymbol{a}_0(\Omega)\right\}
    ,
\end{align}
Since the data symbols in different WDM channels are uncorrelated, we can write \eqref{nli.multi.channel} as
\begin{align}
    \label{main.result}
    \sigma_{\text{NLI},n}^2=\sigma_{\text{SCI}}^2+\sum_{\substack{j=1, j\neq n}}^{N}\sigma_{\text{XPM}}^2(\Omega), \quad \Omega=|j-n|2\pi\Delta f,
\end{align}
where $\Delta f$ is the channel spacing. The SCI and XPM variances given in \eqref{main.result} are expressed as
\begin{align}
    \label{main.result.x}
    \sigma_{\text{SCI}}^2=\sigma_{\text{SCI} ,\text{x}}^2+\sigma_{\text{SCI}, \text{y}}^2,
\end{align}
and
\begin{align}
    \label{XPM.T}
      \sigma_{\text{XPM}}^2(\Omega)=\sigma_{\text{XPM} ,\text{x}}^2(\Omega)+\sigma_{\text{XPM}, \text{y}}^2(\Omega),
\end{align}
resp., in which $\sigma_{\text{SCI}, \text{x}}^2$ and $\sigma_{\text{XPM}, \text{x}}^2$ are the SCI and XPM variances in the x-polarization, resp. The same is true for the y-polarized terms given in \eqref{main.result.x} and \eqref{XPM.T}. The terms $\sigma_{\text{SCI} ,\text{x}}^2$ and $\sigma_{\text{XPM},\text{x}}^2(\Omega)$, given in \eqref{main.result.x} and \eqref{XPM.T}, resp., are equal to
\begin{align}
    \label{total.sci}
    \sigma_{\text{SCI},\text{x}}^{2}=\frac{8}{81}\gamma^2P_{\text{\tiny COI}}^3\left(\Psi_1 S_1+\Psi_2 X_1+\Psi_3 X_2+3Z_1\right),
\end{align}
and
\begin{align}
    \label{final.variance.total.main}
    &\sigma_{\text{XPM},\text{x}}^2(\Omega)=\frac{8}{81}\gamma^2P_{\text{\tiny COI}}P_{\text{\tiny INT}}^2(\Omega)\left(\Phi_1(\Omega)X(\Omega)+6Z(\Omega)\right).
\end{align}
The terms $S_1$, $X_1$, $X_2$, $Z_1$, $X(\Omega)$, and $Z(\Omega)$ in Table~\ref{terms.identical} depend on the spectral properties of the signal, in contrast with  $\Psi_1$, $\Psi_2$, $\Psi_3$ and $\Phi_1$, given in Table~\ref{kerr.terms}, which depend on the modulation format. The SCI and XPM variances in the y-polarization can be obtained from \eqref{total.sci} and  \eqref{final.variance.total.main}, resp., by swapping x and y in \eqref{total.sci}, \eqref{final.variance.total.main} and Table~\ref{varphi.terms}. 

In the special case of independent polarizations that the same format is used in both polarizations, Table~\ref{varphi.terms} yields $\varphi_4=\varphi_3=\varphi_2$ and $\varphi_7=\varphi_5=1$. These values used in Table~\ref{kerr.terms} give $\Psi_1=\varphi_1-9\varphi_2+12$, $\Psi_2=5\varphi_2-10$, $\Psi_3=\varphi_2-2$, and  $\Phi_1=5\varphi_6-10$. These values used in combination with the integral expressions in Table~\ref{terms.identical} can be shown to coincide with the EGN model.
\begin{table}[t]
     \renewcommand{\arraystretch}{1.5}
    \footnotesize
    \centering

      \caption{Integral expressions for the terms used in \eqref{total.sci} and \eqref{final.variance.total.main}. The functions $\rho_{\text{s}}(\cdot)$ and $\rho_{\text{xp}}(\cdot)$ are given in \eqref{rho.s} and \eqref{rho}, resp.}
    \label{terms.identical}

    \medskip

    \begin{tabular}{|@{}c@{~}|@{~}l@{}|}
    \hline
    
    \hline
    Term & Integral Expression \\
    \hline
    \hline
    $S_1$ & 
  $\!\!\frac{1}{T}\!\!\!\int\!\!\frac{\text{d}^3w}{(2\pi)^3} \frac{\text{d}^2w'}{(2\pi)^2}\rho_{\text{s}}(w_1,w_2,w_3)\rho_{\text{s}}^*(w_1',w_2',w_1\!+\!w_3\!+\!w_2'\!-\!w_2\!-\!w_1')$
    \\
    \hline
    $X_1$ & 
    $ \frac{1}{T^2}\int\frac{\text{d}^3w}{(2\pi)^3} \frac{\text{d}w_2'}{2\pi}\rho_{\text{s}}(w_1,w_2,w_3)\rho_{\text{s}}^*(w_1,w_2',w_2'-w_2+w_3)$
    \\
    \hline
    $X_2$ &
$ \frac{1}{T^2}\int\frac{\text{d}^3w}{(2\pi)^3} \frac{\text{d}w_1'}{2\pi}\rho_{\text{s}}(w_1,w_2,w_3)\rho_{\text{s}}^*(w_1',w_2,w_1+w_3-w_1')$
    \\
 \hline
    $Z_1$ &
   $ \frac{1}{T^3}\int\frac{\text{d}^3w}{(2\pi)^3}|\rho_{\text{s}}(w_1,w_2,w_3)|^2$\\
     \hline
     \hline
     
         $X$ & 
    $ \frac{1}{T^2}\int\frac{\text{d}^3w}{(2\pi)^3} \frac{\text{d}w_2'}{2\pi}\rho_{\text{xp}}(w_1,w_2,w_3)\rho_{\text{xp}}^*(w_1-w_2+w_2',w_2',w_3)$
    \\
    \hline
       $Z$ &
   $ \frac{1}{T^3}\int\frac{\text{d}^3w}{(2\pi)^3}|\rho_{\text{xp}}(w_1,w_2,w_3)|^2$\\  
    \hline 
    
    \end{tabular}


\vspace{3mm}
     \caption{The terms used in \eqref{total.sci} and \eqref{final.variance.total.main}. The values of $\varphi_1,\cdots,\varphi_7$ are given in Table~\ref{varphi.terms}.}

    \medskip

   \label{kerr.terms}
   
    \begin{tabular}{|c|l|}
 %
    \hline
    
    \hline
   {Term} &  {Expression}\\
    \hline
    \hline
   $\displaystyle \Psi_1$ &
$\varphi_1-12\varphi_2+24+2\varphi_3+\varphi_4-12\varphi_5$ 
\\
    \hline
    $\displaystyle \Psi_2$ &
$5\varphi_2-15+
5\varphi_5
$
\\
    \hline
    $\displaystyle \Psi_3$ &
$\varphi_2-3+\varphi_5$
\\
    \hline  
\hline
$\displaystyle \Phi_1$ &  $
5\varphi_6-15+
5\varphi_7
$
\\
    \hline
    \end{tabular}
\vspace{3mm}
     \caption{Expressions for the terms $\varphi_1,\cdots,\varphi_7$ used in Table~\ref{kerr.terms}.}

    \medskip

   \label{varphi.terms}
   
    \begin{tabular}{|@{~}c@{~}|@{~}l@{~}|@{~}c@{~}|@{~}l@{~}|@{~}c@{~}|@{~}l@{~}|}
    \hline 
    
    \hline
    Term &  Expression&Term &  Expression&Term &  Expression\\
\hline
\hline
$\varphi_1$ &  $\frac{\mathbb{E}\{|a_{\text{x}}|^6\}}{\mathbb{E}^3\{|a_{\text{x}}|^2\}}$&$\varphi_2$ &  $\frac{\mathbb{E}\{|a_{\text{x}}|^4\}}{\mathbb{E}^2\{|a_{\text{x}}|^2\}}$&$\varphi_3$&  $\frac{\mathbb{E}\{|a_{\text{x}}|^4|a_{\text{y}}|^2\}}{\mathbb{E}^3\{|a_{\text{x}}|^2\}}$\\
\hline
$\varphi_4$ &  $\frac{\mathbb{E}\{|a_{\text{y}}|^4|a_{\text{x}}|^2\}}{\mathbb{E}^3\{|a_{\text{x}}|^2\}}$&  $\varphi_5$&$\frac{\mathbb{E}\{|a_{\text{x}}|^2|a_{\text{y}}|^2\}}{\mathbb{E}^2\{|a_{\text{x}}|^2\}}$&{$\varphi_6$}&{$\frac{\mathbb{E}\{|b_{\text{x}}|^4\}}{\mathbb{E}^2\{|b_{\text{x}}|^2\}}$}\\
\hline
{$\varphi_7$}&{$\frac{\mathbb{E}\{|b_{\text{x}}|^2|b_{\text{y}}|^2\}}{\mathbb{E}^{2}\{|b_{\text{x}}|^2\}}$}&\multicolumn{4}{c|}{} \\
\hline
    \end{tabular}
\end{table}
\begin{table*}[htp]
\caption{The value $\Phi_1$ for 4D constellations chosen from \cite{Erik.database} along with three new constellations proposed in \cite{Chen2019,chen2020analysis,Kojima2017}.
} 
\centering 
\begin{tabular}{l l l l l l l l l l} 
\hline

\hline 
\scriptsize{\bf Modulation} & \scriptsize{\bf $\Phi_1$} & \scriptsize{\bf Modulation} & \scriptsize{\bf $\Phi_1$}&
\scriptsize{\bf Modulation} & \scriptsize{\bf $\Phi_1$}&
\scriptsize{\bf Modulation} & \scriptsize{\bf $\Phi_1$}&
\scriptsize{\bf Modulation} & \scriptsize{\bf $\Phi_1$} \\
\hline 
\scriptsize{biortho4_8} & \scriptsize{$-5$} & 
\scriptsize{tetra4_9} & \scriptsize{$-3.75$} & 
\scriptsize{PM-QPSK} & \scriptsize{$-5$} & \scriptsize{SO-PM-QPSK}& \scriptsize{$-3$} &
 \scriptsize{dicyclic4_24} & \scriptsize{$-5$}\\ 
 \scriptsize{24cell4_24}& \scriptsize{$-5$}& 
 \scriptsize{l4_25} & \scriptsize{$-4.58$} & 
 \scriptsize{b4_32} & \scriptsize{$-4.38$} & \scriptsize{w4_40}& \scriptsize{$-4.05$} &
 \scriptsize{w4_49} & \scriptsize{$-3.65$} \\ 
\scriptsize{b4_64}& \scriptsize{$-4.14$} &
\scriptsize{4D-2A8PSK}\cite{Kojima2017}&
\scriptsize{$-5$} & \scriptsize{4D-64PRS}\cite{Chen2019}&
\scriptsize{$-5$}& 
 \scriptsize{w4_88} & \scriptsize{$-3.87$} &
\scriptsize{4D-OS128}\cite{chen2020analysis}&
\scriptsize{$-3.02$}\\
 \scriptsize{SP-QAM4_128} & \scriptsize{$-3.4$}&
 \scriptsize{w4_145}& \scriptsize{$-3.99$}&
\scriptsize{w4_152} & \scriptsize{$-3.77$}& \scriptsize{w4_169}& \scriptsize{$-3.88$} &
 \scriptsize{PM-16QAM} & \scriptsize{$-3.4$} \\ 
 \scriptsize{w4_256} & \scriptsize{$-3.8$} &  \scriptsize{w4_313}& \scriptsize{$-3.75$} &
 \scriptsize{w4_409} & \scriptsize{$-3.77$} & \scriptsize{w4_464}& \scriptsize{$-3.74$} &
 \scriptsize{cross4_512} & \scriptsize{$-3.57$}\\ \scriptsize{sphere4_512}& \scriptsize{$-3.8$} &
 \scriptsize{SP-cross4_512}& \scriptsize{$-3.45$} &
 \scriptsize{120cell4_600}& \scriptsize{$-5$}&
 \scriptsize{w4_601} & \scriptsize{$-3.81$} & \scriptsize{w4_656}& \scriptsize{$-3.76$} \\
 \scriptsize{w4_800} & \scriptsize{$-3.77$} & \scriptsize{cross4_2048}& \scriptsize{$-3.51$}&
 \scriptsize{SP-QAM4_2048} & \scriptsize{$-3.09$} &
 \scriptsize{PM-64QAM}& \scriptsize{$-3.09$}\\
\hline 
\end{tabular}
\label{table:nonlin} 
\end{table*}
\section{ Numerical Results}\label{numericalSec}

This section is focused on investigating the NLI of 4D modulation formats from the database \cite{Erik.database}. A coherent transmission link consisting
of 100 km spans of a standard single-mode fiber was simulated.
The following parameters were used: Dispersion coefficient
$D=\text{16.5}$ ps/nm/km, nonlinear coefficient $\gamma= \text{1.3}$ 1/W/km,
attenuation $\alpha=\text{0.2}$ dB/km, EDFA noise figure 5 dB, optical center
wavelength 1550 nm, symbol rate $T^{-1}=\text{32}$ Gbaud, and channel spacing $\Delta f=\text{50}$ GHz. 
To compute the experienced NLI of 4D formats, we employ the 4D model defined in \eqref{main.result}--\eqref{final.variance.total.main} together with Tables~\ref{terms.identical} and \ref{kerr.terms}. To relate our work to previously works, we compare our model with the EGN model. 

In this section, we compare 4D constellations in terms of
\begin{align}
    \label{eta_n}
    \eta_n=\frac{\sigma_{\text{NLI},n}^2}{P^3},
\end{align}
assuming $P_{\text{\tiny{COI}}}=P_{\text{\tiny{INT}}}=P$, where $\sigma_{\text{NLI},n}^2$ is defined in \eqref{main.result}. The SNR of the COI $n$ is $\text{SNR}_n = P/(\sigma_{\text{ASE}}^2+\sigma_{\text{NLI},n}^2)$, where $\sigma_{\text{ASE}}^2$ is the variance of the amplified spontaneous emission noise (ASE). 
We first validate the 4D model using the split-step Fourier method (SSFM), and then compare a wide range of 4D constellations. 

\subsection{SSFM Simulations}

Numerically solving
the Manakov equation \eqref{Manakov} for the entire C band is a big challenge. Part of the problem is, of course, high memory
requirements, in addition to the excessive use of very large fast Fourier
transforms.
For this reason, the SSFM study was restricted to a bandwidth of 0.5 THz.
To validate $\eta_{n}$, given in \eqref{eta_n}, ASE-noise-free
SSFM numerical simulations were performed. In the absence of other noise sources, $\eta_{n}$ can be estimated via
the received SNR for each channel $n$
via the relationship
\begin{align}\label{eta.ssfm}
    \eta_{n}\approx\frac{1}{\text{SNR}_n^{\text{est}}P^2}.
\end{align}
The approximate equality in \eqref{eta.ssfm} is due to the fact that the SSFM-based $\text{SNR}^{\text{est}}$ estimates also contain higher order perturbation terms. The SNR for a constellation with $M$
symbols was estimated via
\begin{align}\label{SNR.ssfm}
\text{SNR}_n^{\text{est}}=\frac{\sum_{i=1}^M|\bar{y_i}|^2}{\sum_{i=1}^M\mathbb{E}\{|Y-\bar{y}_i|^2|X=x_i\}},
\end{align}
where $X$ and $Y$ are the random variables representing the
transmitted and received symbols, resp., $x_i$ is the $i$-th constellation
point, and $\bar{y}_i = \mathbb{E}\{Y |X = x_i\}$. A total number of 30000 symbols were used, of which the first 1500 and the last 1500 symbols were removed from the transmitted and received sequences. 
All channels used a flat launch power of $P=0$ dBm.

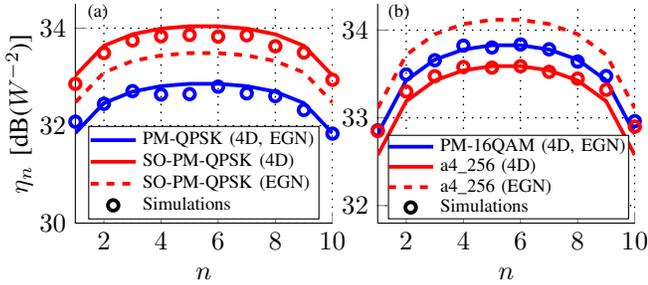
\begin{figure}[!t]
\pgfplotsset{compat=1.3}
\begin{center}
\begin{tikzpicture}[x=0.2cm,y=0.2cm,scale=1,spy using outlines={rectangle,lens={scale=3}, size=2.7cm, connect spies}]
\begin{groupplot}[group style={
group name=my plots, group size=2 by 1,horizontal sep=0.6cm,vertical sep=32pt},
grid,
 grid style = {
    dash pattern = on 0.05mm off 1mm,
    line cap = round,
    black,
    line width = 0.2mm
  },
 ylabel style={yshift=-0.15cm},
axis y line*=left,
axis x line*=bottom,
width=5cm,height=4.5cm]
\nextgroupplot
[
ymin=30,ymax=34.5, 
xmin=1,xmax=10,
xlabel={$n$}, 
legend style={at={(.05,0.23)},anchor=west,font=\scriptsize, row sep=-0.07cm, inner sep=0cm},
legend cell align={left},
legend columns=1,
legend entries={{PM-QPSK (4D, EGN)}, SO-PM-QPSK (4D), SO-PM-QPSK (EGN), Simulations}, 
ylabel={$\eta_{n}$ [dB($W^{-2}$)]},
]
\addlegendimage{no markers,blue,line width=0.5mm,}
\addlegendimage{no markers,red,line width=0.5mm,}
\addlegendimage{no markers,red,dashed, line width=0.5mm,}
\addlegendimage{black,only marks,solid, mark=o,line width=0.5mm,}

\addplot [blue, line width=0.5mm]table[x index=0,y index=1] {Dat/NLIPMQPSKSSFM.dat};

\addplot [red, line width=0.5mm]table[x index=0,y index=1] {Dat/NLISOPMQPSKSSFM.dat};

\addplot [red,dashed, line width=0.5mm]table[x index=0,y index=1] {Dat/NLISOPMQPSKSSFMEGN.dat};

\addplot [red, line width=0.5mm,only marks, mark=o,fill=white]table[x index=0,y index=1] {Dat/NLISOPMQPSKSSFMR.dat};

\addplot [blue, line width=0.5mm, only marks, mark=o,fill=white]table[x index=0,y index=1] {Dat/NLI_PMQPSK_SSFM.dat};
\nextgroupplot
[
ymin=31.8,ymax=34.3, 
xmin=1,xmax=10,
xlabel={$n$}, 
legend style={at={(.04,0.21)},anchor=west,font=\scriptsize, row sep=-0.09cm, inner sep=-0.03cm},
legend cell align={left},
legend columns=1,
legend entries={{PM-16QAM (4D, EGN)},a4_256 (4D),a4_256 (EGN),Simulations},
ytick={32,33,34},yticklabels={
$32$,
$33$,
$34$
}, 
]
\addlegendimage{no markers,blue,line width=0.5mm,}
\addlegendimage{no markers,red,line width=0.5mm,}
\addlegendimage{no markers,red,dashed,line width=0.5mm,}
\addlegendimage{black,only marks,solid, mark=o,line width=0.5mm,}

\addplot [blue, line width=0.5mm, solid]table[x index=0,y index=1] {Dat/NLI16QAMSSFM.dat};

\addplot [red, line width=0.5mm]table[x index=0,y index=1] {Dat/NLIa4256SSFM.dat};

\addplot [red,dashed, line width=0.5mm]table[x index=0,y index=1] {Dat/NLIa4256EGN.dat};

\addplot [blue, line width=0.5mm, only marks, mark=o,fill=white]table[x index=0,y index=1] {Dat/NLI16QAMSSFMR.dat};

\addplot [red, line width=0.5mm,only marks, mark=o,fill=white]table[x index=0,y index=1] {Dat/NLIa4256SSFMRFin.dat};

\end{groupplot}
\node at (1.5,14){\scriptsize(a)}; 
\node at (1.5+20,14){\scriptsize(b)}; 
\end{tikzpicture}
\end{center}
\setlength{\belowcaptionskip}{-12pt}
\vspace*{-2mm}
\caption{$\eta_{n}$ as a function of channel number $n$ after 5 spans. The link consisting of 5 spans supports $N=10$ WDM channels. 
} 
\label{nlc_fig.ssfm}
\end{figure}

A WDM system with $N=10$ channels and four modulation formats, namely PM-QPSK, subset optimized PM-QPSK (SO-PM-QPSK) \cite{Sjodin2013,Erik.database}, PM-16QAM and a4_256 \cite{tobias2015,Erik.database} was simulated. Fig.~\ref{nlc_fig.ssfm} shows the simulation results for $\eta_{n}$ in $\text{dB}(W^{-2})=10\log_{10}(\eta_n\cdot 1W^2)$ using markers for a transmission distance of 500 km.
Fig.~\ref{nlc_fig.ssfm} (a) indicates that the 4D model results for SO-PM-QPSK perfectly follows the simulations, whereas the EGN model fails to estimate the NLI of this format. Fig.~\ref{nlc_fig.ssfm} (b) also illustrates that the results obtained from the 4D model for a4_256 are in good agreement with simulations, while the EGN model results depart from simulations. These results show that the EGN model is inaccurate for the study of arbitrary 4D constellations, and that the NLI can be underestimated (SO-PM-QPSK) or overestimated (a4_256). The proposed 4D model instead has the capacity to predict the NLI of 4D formats with a good level of accuracy. The discrepancy between simulations and the results obtained from the 4D model is on average about 0.2 dB. For PM-QPSK and PM-16QAM, both the EGN model and 4D model give the same results that match the simulation results. In the following section, we attempt to identify the reasons behind an increase or decrease in the NLI estimated from the EGN model and 4D model.

\subsection{Comparing a wide range of constellations}

This section investigates the NLI of 4D constellations propagated in a C-band system. We assume that the entire spectrum is populated with $N=80$ WDM channels, and that the link comprises 10 spans.  
\begin{figure}[htp]
\pgfplotsset{compat=1.3}
\begin{center}
\begin{tikzpicture}[x=0.2cm,y=0.2cm,scale=1]
\begin{groupplot}[group style={
group name=my plots, group size=2 by 2,horizontal sep=1.2cm,vertical sep=72pt},
grid,
 grid style = {
    dash pattern = on 0.05mm off 1mm,
    line cap = round,
    black,
    line width = 0.2mm
  },
 ylabel style={yshift=-0.15cm},
axis y line*=left,
axis x line*=bottom,
width=4.75cm,height=4.25cm]

\nextgroupplot
[
ymin=36,ymax=41, 
xmin=1,xmax=80,
xlabel={$n$}, 
ylabel={$\eta_n$ [dB($W^{-2}$)]},
]

\addplot [blue, line width=0.5mm, solid]table[x index=0,y index=1] {Dat/Total_NLI_PMQPSK_10span.dat};

\addplot [mark=triangle*, green!60!black, fill=white, mark size=3pt, only marks, mark repeat=11]table[x index=0,y index=1] {Dat/Total_NLI_PMQPSK_10span.dat};

\addplot [red, line width=0.5mm]table[x index=0,y index=1] {Dat/Total_NLI_SOPMQPSK_10span.dat};

\addplot [green, line width=0.5mm, dashed]table[x index=0,y index=1] {Dat/Total_NLI_dicyclic_egn.dat};

\addplot [red, line width=0.5mm, dashed]table[x index=0,y index=1] {Dat/Total_NLI_SO_PM_QPSK_EGN.dat};








\nextgroupplot
[
 legend style={at={(-1.1,-0.65)},anchor=west,font=\scriptsize},
legend columns=2,
legend cell align={left},
legend entries={{PM-QPSK (4D, EGN)}, dicyclic4_16 (4D),SO-PM-QPSK (4D), dicyclic4_16 (EGN), SO-PM-QPSK (EGN)},  
ymin=13.5,ymax=17.5, 
xmin=21,xmax=36,
xlabel={Launch Power [dBm]}, 
ylabel={$\text{SNR}_{40}$ [dB]},
xtick={1,6,11,16,21,26,31,36,41},xticklabels={
$-15$,
$-12.5$,
$-10$,
$-7.5$,
$-5$,
$-2.5$,
$0$, 
$2.5$,
$5$, 
}, 
]

\addplot [blue, line width=0.5mm, solid]table[x index=0,y index=1] {Dat/SNR_PMQPSK.dat};

\addplot [green!60!black, only marks, mark=triangle*, fill=white, mark size=3pt, mark repeat=2]table[x index=0,y index=1] {Dat/SNR_PMQPSK.dat};




\addplot [red, line width=0.5mm]table[x index=0,y index=1] {Dat/SNR_SOPMQPSK.dat};

\addplot [green, line width=0.5mm, dashed]table[x index=0,y index=1] {Dat/SNR_dicyclic_egn.dat};

\addplot [red, line width=0.5mm, dashed]table[x index=0,y index=1] {Dat/SNR_sopmqpsk_egn.dat};




\nextgroupplot
[
legend style={at={(-0.03,-0.51)},anchor=west,font=\scriptsize},
legend cell align={left},
legend columns=3,
legend entries={{PM-16QAM (4D, EGN)}, a4_256 (4D), a4_256 (EGN)}, 
ymin=37,ymax=39.8, 
xmin=1,xmax=80,
xlabel={$n$}, 
ylabel={$\eta_n$ [dB($W^{-2}$)]},
]
\addplot [blue, line width=0.5mm, solid]table[x index=0,y index=1] {Dat/Total_NLI_PM16QAM_10span.dat};

\addplot [red, line width=0.5mm, solid]table[x index=0,y index=1] {Dat/Total_NLI_256D4_10span.dat};



\addplot [red, line width=0.5mm, dashed]table[x index=0,y index=1] {Dat/Total_NLI_a4_256_egn.dat};

\nextgroupplot
[
ymin=15,ymax=17.5, 
xmin=25,xmax=35,
xlabel={Launch Power [dBm]}, 
ylabel={$\text{SNR}_{40}$ [dB]},
xtick={1,6,11,16,21,26,31,36,41},xticklabels={
$-15$,
$-12.5$,
$-10$,
$-7.5$,
$-5$,
$-2.5$,
$0$, 
$2.5$,
$5$, 
}, 
]

\addplot [blue, line width=0.5mm, solid]table[x index=0,y index=1] {Dat/SNR_PM_16QAM.dat};

\addplot [red, line width=0.5mm, solid]table[x index=0,y index=1] {Dat/SNR_a4_256.dat};



\addplot [red, line width=0.5mm, dashed]table[x index=0,y index=1] {Dat/SNR_a4_256_egn.dat};




\end{groupplot}
\draw[->,>=stealth,thin] (60.5-18.5-22.5-11.4+4-3.5,20.5-6.8-9+1.1+2.5-3.4) -- (60.5-18.5-22.5-11.4+4-3.5,26-2.8-9+0.5+5.7-8.2);
\node at (58-18.75-22.5-11.4+5.2-1+4.15-3,23-5.3+1.7-9+0.5+1.2+4-6) {\footnotesize{$2.8$} \scriptsize{dB}};

\draw[->,>=stealth,thin] (60.5-18.5-22.5-11.4-0.5+3.5-3.4,20.5-6.8-9+1.1+2.3-3) -- (60.5-18.5-22.5-11.4-0.5+3.5-3.4,26-2.8-9+0.5-4.8+3.85-5.4);
\node at (58-18.75-22.5-11.4+5.2-1.5-4+1+3.3-4,23-5.3+1.7-9+0.5+1.2-5+5.2-6.35) {\scriptsize{$1.34$} \scriptsize{dB}};



\draw[<-,>=stealth,thin] (60.5-18.5-22.5-11.4-0.5+25.5-16.9+16.5,20.5-6.8-9+1.1+4.5-23.5+21.8) -- (60.5-18.5-22.5-11.4-0.5+25.5-16.9+16.5,26-2.8-9+0.5-4.8+4.75-22+19.5);
\node at (58-18.75-22.5-11.4+5.2-1.5-4+25.5+1.3-16.9+15.5,23-5.3+1.7-9+0.5+1.2-5.3+5-23+18.2) {\scriptsize{$1.1$} \scriptsize{dB}};
\draw [->] (60.5-18.5-22.5-11.4-0.5+25.1,26-2.8-9+0.5-4.8) arc (-95:-9:-13pt);

\node at (1.3,12.7) {\scriptsize{(a)}};
\node at (1.3+22,12.7) {\scriptsize{(b)}};
\node at (1.3,12.7-26.5) {\scriptsize{(c)}};
\node at (1.3+22,12.7-26.5) {\scriptsize{(d)}};
\end{tikzpicture}
\end{center}
\setlength{\belowcaptionskip}{-12pt}
\vspace*{-2mm}
\caption{(a) and (c) illustrate $\eta_n$, defined in \eqref{eta_n},
as a function of channel number $n$ after 10 spans, while (b) and (d) illustrate the SNR of COI, i.e.,  $\text{SNR}_{40}$, as a function of launch power after 10 spans. The full C-band spectrum can accommodate $N=80$ WDM channels. 
} 
\label{nlc_fig4}
\end{figure}
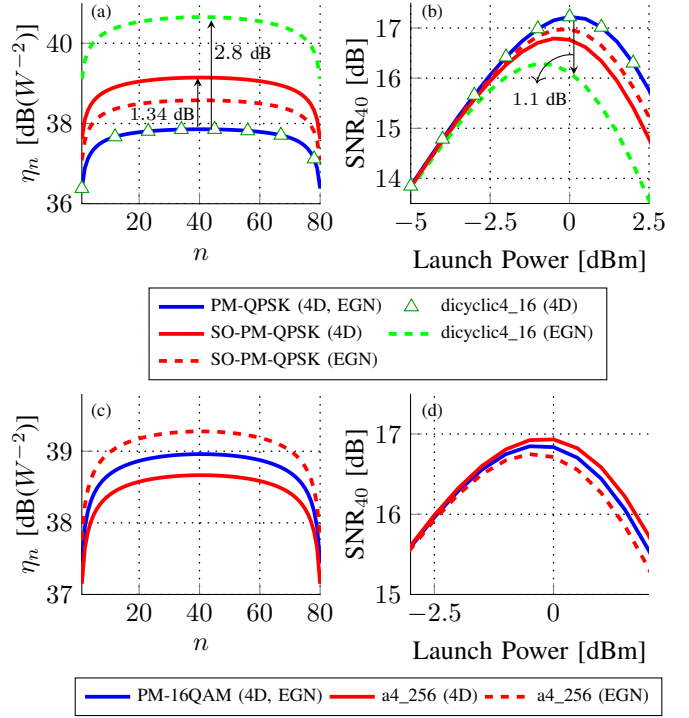
Figs.~\ref{nlc_fig4} (a) and (c) compare different formats in terms of 
$\eta_n$,
 while Figs.~\ref{nlc_fig4} (b) and (d) compare them in terms of $\text{SNR}_{40}$. We interpret the first two coordinates in a coordinate list of \cite{Erik.database} as the x polarization and the last two as the y polarization.
 Figs.~\ref{nlc_fig4} (a) and (b) give information about PM-QPSK, SO-PM-QPSK and dicyclic4_16 \cite{zetterberg1977codes,Erik.database}.
    We benchmark the 4D model against the EGN model in this figure.
As can be seen in Fig.~\ref{nlc_fig4} (a), the curves are highest in the middle of spectrum. 
The 4D model indicates that over the entire spectrum shown in Fig.~\ref{nlc_fig4} (a), SO-PM-QPSK undergoes the most NLI, while PM-QPSK and dicyclic4_16 experience the least. It is also noticeable that PM-QPSK and dicyclic4_16 have the same NLI.
The difference between the experienced NLI for SO-PM-QPSK and PM-QPSK is about 1.34 dB. This means that SO-PM-QPSK is more vulnerable to the Kerr nonlinearity than PM-QPSK.
The SCI and XPM terms are responsible for this gap. The impact of SCI on the COI is high, but the XPM effects in multi-channel WDM systems are even higher, and therefore, the better part of this deviation stems form the XPM terms. To be more specific, the origin of this discrepancy comes form the fact that  $\Phi_1$, given in Table~\ref{kerr.terms}, for SO-PM-QPSK ($\Phi_1=-3$) is larger than for PM-QPSK ($\Phi_1=-5$).

{From the curve with triangles (4D model) to the green curve (EGN model), there is a 2.8 dB increase in the NLI for dicyclic4_16, with SNR falling by around 1.1 dB to approximately 16.1 dB (see Fig.~\ref{nlc_fig4} (b)).
This implies that the EGN model significantly overestimates the NLI for dicyclic4_16. 
This is because $\varphi_7=0$ for this format according to Table~\ref{varphi.terms}, whereas the EGN model corresponds to setting $\varphi_7=1$ for any format.
On the other hand, we can see that the EGN model underestimates the NLI of SO-PM-QPSK in comparison with the 4D model. This is because the term $\varphi_7$ is lower for the EGN model ($\varphi_7=1$) than for the 4D model ($\varphi_7=1.2$).}


Fig.~\ref{nlc_fig4} (c) and (d) compare the PM-16QAM and a4_256 formats in terms of $\eta_n$ and SNR, resp. Fig.~\ref{nlc_fig4} (c) shows that PM-16QAM is at a disadvantage compared with a4_256. The deviation of the NLI between the PM-16QAM and a4_256 formats, as shown in Fig.~\ref{nlc_fig4} (c), is about 0.3 dB. 
This deviation may be rooted in the value of $\Phi_1$ which is smaller for a4_256 ($\Phi_1=-3.8$) than for PM-16QAM ($\Phi_1=-3.4$).
We can also see in Fig.~\ref{nlc_fig4}~(c) that the EGN model overestimates the NLI of a4_256 by about 0.6 dB.
It is clear from Fig.~\ref{nlc_fig4}~(d) that the SNR for a4_256 falls from about 17 dB (4D model) to around 16.8 dB (EGN model) at 0 dBm launch power.

As mentioned earlier, the experienced amount of NLI is dependent on $\Psi_1,\Psi_2,\Psi_3$, and $\Phi_1$ given in Table~\ref{kerr.terms}, none of which is more important than $\Phi_1$. Table~\ref{table:nonlin} shows the value of $\Phi_1$ for constellations selected from \cite{Erik.database} as well as three constellations proposed in \cite{Chen2019,chen2020analysis,Kojima2017}. All the constellations shown in Table~\ref{table:nonlin} satisfy the assumptions made in Sec.~\ref{key.results}. It is clear that SO-PM-QPSK and four-dimensional orthant-symmetric
128-ary modulation (4D-OS128) \cite{chen2020analysis} generate higher NLI than do other formats. For SO-PM-QPSK and 4D-OS128, $\Phi_1$ is around $-3$, which is higher than the others. The lowest amount of NLI belongs to constellations whose $\Phi_1$ equals to $-5$, meaning that these constellations undergo approximately the same NLI as PM-QPSK.

\section{Conclusion}\label{conclusion}

A nonlinear model which analytically models the impact of the Kerr nonlinearity on a 4D signal space was proposed and
analyzed in detail. The model applies to zero-mean dual-polarization 4D formats which are symmetric with respect to the origin and have equal energy on the two polarization components. Unlike the GN and EGN models, we consider the interpolarization dependency so as to derive a 4D nonlinear model. The proposed model accounts for the SCI and XPM nonlinear terms. This is because the SCI and XPM are the predominant nonlinear
terms in multi-channel WDM systems. We have compared different 4D modulation formats in terms of the experienced NLI, and showed that the derived model is a powerful tool to find 4D formats which are more resistant to the NLI.

\appendix

\section{Nonlinear analysis}\label{NLI.analysis}
\vspace{-1mm}
Our intention, in this section, is to study the variance of SCI (the first term on the right-hand side of \eqref{first.order.solution}) and XPM (the second term on the right-hand side of \eqref{first.order.solution}) terms. We proceed by computing the variance of SCI. 
\subsection{SCI variance}
For the sake of brevity, we only focus on the x-polarized element of \eqref{first.order.solution} because we can obtain the results for the y-polarized component under the substitution x$\rightarrow$y, y$\rightarrow$x. 
Using \eqref{first.order.solution}, the x-polarized component of SCI term is given by
\begin{align}
    \label{nlin.S}
   \! \Delta a_{0, \text{SCI}, \text{x}}\!=&i\frac{8}{9}\gamma\sum_{h,k,l}S_{h,k,l}\Big(a_{h,\text{x}}a_{k,\text{x}}^*a_{l,\text{x}}+a_{h,\text{y}}a_{k,\text{y}}^*a_{l,\text{x}}\Big).
\end{align}
The variance of \eqref{nlin.S} is therefore equal to
\begin{align}
\label{nli.variance.sci}
\sigma_{\text{SCI} ,\text{x}}^2=\mathbb{E}\{\Delta a_{0,\text{SCI}, \text{x}}\Delta a_{0,\text{SCI}, \text{x}}^*\}-\mathbb{E}\{\Delta a_{0,\text{SCI}, \text{x}}\}\mathbb{E}\{\Delta a_{0,\text{SCI}, \text{x}}^*\},
\end{align}
where $\mathbb{E}\{\Delta a_{0,\text{SCI}, \text{x}}\}=0$. This is because under the assumptions made in Sec.~\ref{key.results}, we have $\mathbb{E}\{a_{h,\text{x}}\}=  \mathbb{E}\{a_{h,\text{x}}^2\}= \mathbb{E}\{a_{h,\text{x}}a_{h,\text{y}}^*\}=\mathbb{E}\{|a_{h,\text{x}}|^2a_{h,\text{x}}\}=\mathbb{E}\{|a_{h,\text{y}}|^2a_{h,\text{x}}\}=0$
 (see \cite[Appendix~A]{golani2016modeling}), and as a result, $\mathbb{E}\{a_{h,\text{x}}a_{k,\text{x}}^*a_{l,\text{x}}\}=\mathbb{E}\{a_{h,\text{y}}a_{k,\text{y}}^*a_{l,\text{x}}\}=0$ for all $h$, $k$, and $l$. Substituting \eqref{nlin.S} into \eqref{nli.variance.sci} gives
 \begin{align}
\label{nli.variance.sci.1}
&\sigma_{\text{SCI} ,\text{x}}^2=\frac{64}{81}\gamma^2\sum_{h,k,l,h',k',l}S_{h,k,l}S_{h',k',l'}^*\Big(\nonumber\\&\mathbb{E}\{a_{h,\text{x}}a_{k,\text{x}}^*a_{l,\text{x}}a_{h',\text{x}}^*a_{k',\text{x}}a_{l',\text{x}}^*\}+\mathbb{E}\{a_{h,\text{x}}a_{k,\text{x}}^*a_{l,\text{x}}a_{h',\text{y}}^*a_{k',\text{y}}a_{l',\text{x}}^*\}\nonumber\\&\!\!\!+\!
\mathbb{E}\{a_{h,\text{y}}a_{k,\text{y}}^*a_{l,\text{x}}a_{h',\text{x}}^*a_{k',\text{x}}a_{l',\text{x}}^*\}\!+\!\mathbb{E}\{a_{h,\text{y}}a_{k,\text{y}}^*a_{l,\text{x}}a_{h',\text{y}}^*a_{k',\text{y}}a_{l',\text{x}}^*\}\!\Big).
\end{align}
 We can rewrite \eqref{nli.variance.sci.1} as
\begin{align}
\sigma_{\text{SCI},\text{x}}^2 = \sum_{i=1}^4 \sigma_{\text{SCI},\text{x},i}^2,
\end{align}
where $\sigma_{\text{SCI},\text{x},i}^2$ represents the $i$-th term in \eqref{nli.variance.sci.1}. We only give the procedure of calculating $\sigma_{\text{SCI},\text{x},4}^2$ in detail, and we can follow the same approach for the others. The contribution of $\sigma_{\text{SCI},\text{x},1}^2$ was calculated in the first term of Eq.~(36) and  Eq.~(105) of \cite{carena2014accuracy}, and $\sigma_{\text{SCI},\text{x},4}^2$ is more challenging to compute than the second and third terms, which is why we focus on calculating this term.
The term $\sigma_{\text{SCI},\text{x},4}^2$ is given by 
\begin{align}\label{sci.x.4}
\sigma_{\text{SCI},\text{x},4}^2\!\!=\!\!\frac{64}{81}\gamma^2\!\!\!\!\!\!\!\sum_{h,k,l,h',k',l}\!\!\!\!\!\!\!S_{h,k,l}S_{h',k',l'}^* \mathbb{E}\{a_{h,\text{y}}a_{k,\text{y}}^*a_{l,\text{x}}a_{h',\text{y}}^*a_{k',\text{y}}a_{l',\text{x}}^*\},
\end{align}
whose expectation term is equal to\cite[Eqs.~(26) and (27)]{golani2016modeling}
\begin{align}
    \label{6.moment}
&\mathbb{E}\{a_{h,\text{y}}a_{k,\text{y}}^*a_{l,\text{x}}a_{h',\text{y}}^*a_{k',\text{y}}a_{l',\text{x}}^*\}=\nonumber\\&
\left\{
\begin{array}{rl}
\mathbb{E}\{|a_{\text{x}}|^2|a_{\text{y}}|^4\}, & h=k=l=h'=k'=l',\\
\mathbb{E}\{|a_{\text{y}}|^2\}\mathbb{E}\{|a_{\text{x}}|^2|a_{\text{y}}|^2\}, &  h=h'\neq l=k=k'=l',\\
\mathbb{E}\{|a_{\text{y}}|^2\}\mathbb{E}\{|a_{\text{x}}|^2|a_{\text{y}}|^2\}, &  h=k\neq l=h'=k'=l',\\
\mathbb{E}\{a_{\text{y}}a_{\text{x}}^*\}\mathbb{E}\{|a_{\text{y}}|^2a_{\text{y}}^*a_{\text{x}}\}, &  h=l'\neq l=k=h'=k',\\
\mathbb{E}\{a_{\text{y}}^*a_{\text{x}}\}\mathbb{E}\{|a_{\text{y}}|^2a_{\text{y}}a_{\text{x}}^*\}, &  k=l\neq h=h'=k'=l',\\
\mathbb{E}\{|a_{\text{y}}|^2\}\mathbb{E}\{|a_{\text{x}}|^2|a_{\text{y}}|^2\}, &  k=k'\neq h=l=h'=l',\\
\mathbb{E}\{a_{\text{y}}^*a_{\text{x}}\}\mathbb{E}\{|a_{\text{y}}|^2a_{\text{y}}a_{\text{x}}^*\}, &  l=h'\neq h=k=k'=l',\\
\mathbb{E}\{|a_{\text{x}}|^2\}\mathbb{E}\{|a_{\text{y}}|^4\}, &  l=l'\neq h=k=h'=k',\\
\mathbb{E}\{|a_{\text{y}}|^2\}\mathbb{E}\{|a_{\text{x}}|^2|a_{\text{y}}|^2\}, &  h'=k'\neq h=k=l=l',\\
\mathbb{E}\{a_{\text{y}}a_{\text{x}}^*\}\mathbb{E}\{|a_{\text{y}}|^2a_{\text{y}}^*a_{\text{x}}\}, &  k'=l'\neq h=k=l=h',\\
\mathbb{E}^2\{|a_{\text{y}}|^2\}\mathbb{E}\{|a_{\text{x}}|^2\}, &  h=h'\neq k=k'\neq l=l',\\
\mathbb{E}\{|a_{\text{y}}|^2\}\mathbb{E}\{a_{\text{x}}a_{\text{y}}^*\}\mathbb{E}\{a_{\text{y}}a_{\text{x}}^*\}, &  h=k\neq l=h'\neq k'=l',\\
\mathbb{E}^2\{|a_{\text{y}}|^2\}\mathbb{E}\{|a_{\text{x}}|^2\}, &  h=k\neq l=l'\neq k'=h',\\
\mathbb{E}\{|a_{\text{y}}|^2\}\mathbb{E}\{a_{\text{x}}a_{\text{y}}^*\}\mathbb{E}\{a_{\text{y}}a_{\text{x}}^*\}, &  h=h'\neq k=l\neq k'=l',\\
\mathbb{E}\{|a_{\text{y}}|^2\}\mathbb{E}\{a_{\text{x}}a_{\text{y}}^*\}\mathbb{E}\{a_{\text{y}}a_{\text{x}}^*\}, &  h=l'\neq k=l\neq h'=k',\\
\mathbb{E}\{|a_{\text{y}}|^2\}\mathbb{E}\{a_{\text{x}}a_{\text{y}}^*\}\mathbb{E}\{a_{\text{y}}a_{\text{x}}^*\}, &  h=l'\neq k=k'\neq l=h'.\\
\end{array}\right.
\end{align}
We can hence write \eqref{sci.x.4} as
\begin{align}\label{sci.x.4.expansion}
\sigma_{\text{SCI},\text{x},4}^2=\sum_{j=1}^{16}\sigma_{\text{SCI},\text{x},4,j}^2,
\end{align}
where $\sigma_{\text{SCI},\text{x},4,j}^2$ stands for the contribution of the $j$-th case, given in \eqref{6.moment}, to \eqref{sci.x.4}. We first remove from \eqref{6.moment} the terms which involve $\mathbb{E}\{a_{\text{x}}a_{\text{y}}^*\}$ or $\mathbb{E}\{a_{\text{x}}^*a_{\text{y}}\}$. By substituting \eqref{S.coef.alt.freq} into \eqref{sci.x.4}, we can write $\sigma_{\text{SCI},\text{x},4,1}^2$, given in \eqref{sci.x.4.expansion}, as
\begin{align}\label{first.intra.x.contribution}
   & \sigma_{\text{SCI},\text{x},4,1}^{2}=\frac{64}{81}\gamma^2\mathbb{E}\{|a_{\text{x}}|^2|a_{\text{y}}|^4\}\int \frac{\text{d}^3w}{(2\pi)^3} \frac{\text{d}^3w'}{(2\pi)^3} \rho_{\text{s}}(w_1,w_2,w_3)\nonumber\\&\cdot\rho_{\text{s}}^*(w_1',w_2',w_3')\sum_{h} \text{e}^{i(w_1-w_2+w_3-w_1'+w_2'-w_3')hT}.
\end{align}
Using the identity \cite[Eq.~(14)]{dar2015inter}
\begin{align}\label{identity.delta}
    \sum_{k=-\infty}^{\infty}\text{e}^{ikT{w_1}}=\frac{2\pi}{T}\sum_{n=-\infty}^{\infty}\delta({w_1}-\frac{2\pi n}{T}),
    \end{align}
for the sinc pulse and considering \eqref{half.power}, we can write \eqref{first.intra.x.contribution} as 
\begin{align}\label{cont.sixth.order.moment}
  &\sigma_{\text{SCI},\text{x},4,1}^{2}=\frac{8}{81}\gamma^2P_{\text{\tiny COI}}^3\varphi_4 S_1,
\end{align}
where $S_1$ is given in Table~\ref{terms.identical} and $\varphi_4$ is given in Table~\ref{varphi.terms}.
%

\vspace{-1mm}
By using \eqref{S.coef.alt.freq} once again in \eqref{sci.x.4}, and considering \eqref{half.power} the term $\sigma_{\text{SCI},\text{x}, 4,2}^2$, given in \eqref{sci.x.4.expansion}, is equal to
\begin{align}
    \label{second.case.contribution}
   &\sigma_{\text{SCI},\text{x},4,2}^{2}=\frac{8}{81}\gamma^2P_{\text{\tiny COI}}^3\varphi_5\int\frac{\text{d}^3w}{(2\pi)^3}\frac{\text{d}^3w'}{(2\pi)^3} \rho_{\text{s}}(w_1,w_2,w_3)\nonumber\\&\cdot\rho_{\text{s}}^*(w_1',w_2',w_3')\sum_{h\neq l}\text{e}^{i(w_1-w_1')hT+i(w_3-w_2+w_2'-w_3')lT},
\end{align}
where $\varphi_5$ is given in Table ~\ref{varphi.terms}. Using the same approach given in \cite[Eq.~(29)]{golani2016modeling}, we have
\begin{align}
\label{neq.double.sum}
&\sum_{h\neq l}\text{e}^{i(w_1-w_1')hT}\text{e}^{i(w_3-w_2+w_2'-w_3')lT}=\sum_{h, l}\text{e}^{i(w_1-w_1')hT}\nonumber\\&\cdot\text{e}^{i(w_3-w_2+w_2'-w_3')lT}-\sum_{h}\text{e}^{i(w_1-w_1'+w_3-w_2+w_2'-w_3')hT}.
\end{align}
Considering \eqref{identity.delta}, we can rewrite \eqref{neq.double.sum} for the sinc pulse as
\begin{align}
\label{neq.double.sum.identity}
&\!\sum_{h\neq l}\!\text{e}^{i(w_1-w_1')hT}\!\text{e}^{i(w_3-w_2\!+w_2'\!-w_3')lT}\!\!=\!
\frac{4\pi^2}{T^2}\delta(\!w_3\!-\!w_2\!+\!w_2'\!-\!w_3')\nonumber\\& \cdot\delta(w_1-w_1')-\frac{2\pi}{T}\delta(w_1\!-\!w_1'\!+\!w_3\!-\!w_2\!+\!w_2'\!-\!w_3').
\end{align}
By inserting \eqref{neq.double.sum.identity} into \eqref{second.case.contribution}, we get
\begin{align}
    \label{second.case.contribution.f}
   \sigma_{\text{SCI},\text{x},4,2}^{2}&=\frac{8}{81}\gamma^2P_{\text{\tiny COI}}^3\varphi_5\left(X_1-S_1\right),
   \end{align}
 where $X_1$ and $S_1$ are given in Table~\ref{terms.identical}. 
Considering \eqref{S.coef.alt.freq},  \eqref{half.power}, \eqref{sci.x.4}, \eqref{sci.x.4.expansion}, \eqref{neq.double.sum} and \eqref{neq.double.sum.identity}, we can express $\sigma_{\text{SCI},\text{x},4,3}^{2}$, given in \eqref{sci.x.4.expansion}, as
\begin{align}
    \label{third.case.contribution}
  & \sigma_{\text{SCI},\text{x},4,3}^{2}=\frac{8}{81}\gamma^2P_{\text{\tiny COI}}^3\varphi_5\int\frac{\text{d}^3w}{(2\pi)^3}\frac{\text{d}^3w'}{(2\pi)^3} \rho_{\text{s}}(w_1,w_2,w_3)\nonumber\\&\cdot\rho_{\text{s}}^*(w_1',w_2',w_3')\Big(\frac{4\pi^2}{T^2}\delta(w_3-w_1'+w_2'-w_3')\delta(w_1-w_2)\nonumber\\&-\frac{2\pi}{T}\delta(w_1-w_2+w_3-w_1'+w_2'-w_3')\Big).
\end{align}
The term $\delta(w_1-w_2)$ is a bias term and should be discarded.  Bias terms are those for which $w_2=w_1$, $w_2=w_3$, $w_2'=w_1'$, or $w_2'=w_3'$.  These terms create a constant phase shift, and thus, irrelevant for the noise variance we would like to compute (see \cite[Sec.~VIII, Eqs. (63)--(67)]{A.Mecozzi2012}, \cite[Sec.~3, Eq.~(17)]{dar2013properties}, \cite[Appendix A]{carena2014egn}, \cite[Sec.IV-B and the text after (63)]{poggiolini2012detailed} and \cite[Appendix C]{Pontus_JLT_modeling_2013}). Eq.~\eqref{third.case.contribution} is therefore reduced to
\begin{align}
    \label{third.case.contribution.f}
  & \sigma_{\text{SCI},\text{x},4,3}^{2}=-\frac{8}{81}\gamma^2P_{\text{\tiny COI}}^3\varphi_5S_1,
\end{align}
and we can express the same formula for $\sigma_{\text{SCI},\text{x},4,9}^{2}$. 
Following the same approach, the term $\sigma_{\text{SCI},\text{x},4,6}^{2}$, given in \eqref{sci.x.4.expansion}, contributes to \eqref{sci.x.4} as
\begin{align}
    \label{6th.case.contribution.f}
  & \sigma_{\text{SCI},\text{x},4,6}^{2}=\frac{8}{81}\gamma^2P_{\text{\tiny COI}}^3\varphi_5\left(X_2-S_1\right),
\end{align}
where $X_2$ is given in Table~\ref{kerr.terms}.

The last step in calculating  \eqref{sci.x.4.expansion} is to investigate the impact of the last six situations on the NLI variance. We start with $\sigma_{\text{SCI},\text{x},4,11}^{2}$, which contributes to \eqref{sci.x.4.expansion} as
\begin{align}
    \label{sci.11.contribution}
  & \sigma_{\text{SCI},\text{x},4,11}^{2}\!=\!\!\frac{64}{81}\gamma^2\mathbb{E}^2\{|a_{\text{y}}|^2\}\mathbb{E}\{|a_{\text{x}}|^2\}\!\!\!\int\!\!\!\frac{\text{d}^3w}{(2\pi)^3}\frac{\text{d}^3w'}{(2\pi)^3} \rho_{\text{s}}(\!w_1,\!w_2,\!w_3)\nonumber\\&\cdot\rho_{\text{s}}^*(w_1',w_2',w_3')\sum_{h\neq k\neq l}\text{e}^{i(w_1-w_1')hT}\text{e}^{-i(w_2-w_2')kT}\text{e}^{i(w_3-w_3')lT},
\end{align}
where the triple summation is expressed as
\begin{align}
    \label{neq.triple.sum}
    &\sum_{h\neq k\neq l}\!\!\!\text{e}^{i(w_1-w_1')hT-i(w_2-w_2')kT+i(w_3-w_3')lT}\!=\!
    \sum_{h, k, l}\text{e}^{i(w_1-w_1')hT}\nonumber\\&\!\!\!\cdot\text{e}^{-i(w_2-w_2')kT+i(w_3-w_3')lT}\!\!\!-\!\!\!\!\!
       \sum_{h=k\neq l}\!\!\!\!\text{e}^{i(w_1-w_1'-w_2+w_2')hT+i(w_3-w_3')lT}\nonumber\\&-\!\!\!
          \sum_{h=l\neq k}\!\!\!\text{e}^{i(w_1-w_1'+w_3-w_3')hT}\text{e}^{-i(w_2-w_2')kT}-
             \sum_{h\neq k= l}\text{e}^{i(w_1-w_1')hT}\nonumber\\&\cdot\text{e}^{i(-w_2+w_2'+w_3-w_3')kT}+
     2\sum_{h}\text{e}^{i(w_1-w_1'-w_2+w_2'+w_3-w_3')hT}.
\end{align}
Considering \eqref{half.power}, \eqref{identity.delta}, \eqref{neq.triple.sum}, and \eqref{sci.11.contribution}, we have
\begin{align}
    \label{sci.11.contribution.delta}
  & \sigma_{\text{SCI},\text{x},4,11}^{2}=\frac{8}{81}\gamma^2P_{\text{\tiny COI}}^3\int\frac{\text{d}^3w}{(2\pi)^3}\frac{\text{d}^3w'}{(2\pi)^3} \rho_{\text{s}}(w_1,w_2,w_3)\nonumber\\&\cdot\rho_{\text{s}}^*(w_1',w_2',w_3')\Big(\frac{8\pi^3}{T^3}\delta(w_1-w_1')\delta(w_2'-w_2)\delta(w_3-w_3')\nonumber\\&\!-\!
  \frac{4\pi^2}{T^2}\delta(w_1-w_1'+w_2'-w_2)\delta(w_3-w_3')-
  \frac{4\pi^2}{T^2}\delta(w_2'-w_2)\nonumber\\&\cdot\delta(w_1-w_1'+w_3-w_3')-
  \frac{4\pi^2}{T^2}\delta(w_2'-w_2+w_3-w_3')\nonumber\\&\cdot\delta(w_1-w_1')+
  \frac{8\pi}{T}\delta(w_1-w_1'+w_2'-w_2+w_3-w_3')
  \Big),
\end{align}
which can be written as
\begin{align}
    \label{sci.11.contribution.delta.f}
  &\sigma_{\text{SCI},\text{x},4,11}^{2}=\frac{8}{81}\gamma^2P_{\text{\tiny COI}}^3\left(Z_1\!-\!2X_1\!-\!X_2\!+\!2S_1
  \right),
\end{align}
where $Z_1$, $X_1$, $X_2$ and $S_1$ are given in Table~\ref{terms.identical}. The same approach holds for ${\sigma_{\text{SCI},\text{x},4,13}^{2}}$, but the bias terms should not be taken into account. 
Considering \eqref{cont.sixth.order.moment},  \eqref{second.case.contribution.f}, \eqref{third.case.contribution.f}, \eqref{6th.case.contribution.f} and \eqref{sci.11.contribution.delta.f}, and using \eqref{half.power} and \eqref{fourth.order.moment}, we can express \eqref{sci.x.4} as
\begin{align}
    \label{nli.sci.fourth.moment.total}
   \sigma_{\text{SCI},\text{x},4}^{2}\!=&\frac{8}{81}\gamma^2P_{\text{\tiny COI}}^3[(\varphi_4-4\varphi_5-\varphi_2+4) S_1+ \!(\varphi_5\!+\!\varphi_2\!-\!3) X_1\nonumber\\&+ (\varphi_5-1)X_2+Z_1],
\end{align}
which is called the interpolarization nonlinear effect, and the term $\varphi_2$ is given in Table~\ref{varphi.terms}. This expression is not available in the literature. 

The contributions of $\sigma_{\text{SCI},\text{x},1}^{2}$, $\sigma_{\text{SCI},\text{x},2}^{2}$, and $\sigma_{\text{SCI},\text{x},3}^{2}$, given in \eqref{sci.x.4}, can be calculated through the same procedure, so their detailed derivations will not be repeated here, and we only give the final results for them as follows
\begin{align}
    \label{nli.sci.second.moment.total}
  \sigma_{\text{SCI},\text{x},2}^{2}=&\sigma_{\text{SCI},\text{x},3}^{2}=\frac{8}{81}\gamma^2P_{\text{\tiny COI}}^3[(\varphi_3-4\varphi_5-\varphi_2+4) S_1\nonumber\\&+ (2\varphi_5-2)X_1],
\end{align}
\begin{align}
    \label{intra.total.x}
   \sigma_{\text{SCI},\text{x},1}^{2}=&\frac{8}{81}\gamma^2P_{\text{\tiny COI}}^3[(\varphi_1-9\varphi_2+12) S_1+ \left(4\varphi_2-8\right) X_1\nonumber\\&+(\varphi_2-2)X_2+2Z_1],
\end{align}
and we call \eqref{intra.total.x} the intra-polarization nonlinear effect. 
By excluding the bias terms{\footnote{Bias terms in \cite[Eq.~(105)]{carena2014accuracy} are those which involve $\delta_{m-n}$, $\delta_{k-n}$, $\delta_{m'-n'}$ and $\delta_{k'-n'}$.}} from \cite[Eq.~(105)]{carena2014accuracy}, and using it into the first term of \cite[Eq.~(36)]{carena2014accuracy}, we can get \eqref{intra.total.x}.
Putting \eqref{intra.total.x}, \eqref{nli.sci.second.moment.total} and \eqref{nli.sci.fourth.moment.total} together, we
obtain the total variance of the SCI nonlinear term \eqref{nli.variance.sci.1}, which is expressed as \eqref{total.sci} with coefficients from Tables~\ref{terms.identical} and \ref{varphi.terms}.

\vspace{-5mm}
\subsection{XPM variance}

Here we calculate $\sigma_{\text{XPM},\text{x}}^2(\Omega)$ in \eqref{final.variance.total.main} for a single pair of channels with fixed separation $\Omega$. For notational convenience, the dependence on $\Omega$ is dropped throughout the section.
As mentioned in Sec.~\ref{preliminaries}, the second term of \eqref{first.order.solution} gives rise to the XPM nonlinear term. The x-polarized component of this term is  
\begin{align}
    \label{nlin.x}
    i\frac{8}{9}\gamma\sum_{h,k,l}X_{h,k,l}\Big(\!2b_{h,\text{x}}b_{k,\text{x}}^*a_{l,\text{x}}+b_{h,\text{y}}b_{k,\text{y}}^*a_{l,\text{x}}+ b_{h,\text{x}}b_{k,\text{y}}^*a_{l,\text{y}}\Big),
\end{align}
whose variance is equal to
\begin{align}
\label{nli.variance.alt.reduced}
&\sigma_{\text{XPM}, \text{x}}^2=\frac{64}{81}\gamma^2\!\!\!\!\sum_{h,k,l,h',k',l'}\!\!\!X_{h,k,l}X^*_{h',k',l'}\Big(4\mathbb{E}\{b_{h,\text{x}}b_{k,\text{x}}^*b_{h',\text{x}}^*b_{k',\text{x}}\}\nonumber\\&\cdot\mathbb{E}\{a_{l,\text{x}}a_{l',\text{x}}^*\}+
2\mathbb{E}\{b_{h,\text{x}}b_{k,\text{x}}^*b_{h',\text{y}}^*b_{k',\text{y}}\}\mathbb{E}\{a_{l,\text{x}}a_{l',\text{x}}^*\}\nonumber\\&+
2\mathbb{E}\{b_{h,\text{y}}b_{k,\text{y}}^*b_{h',\text{x}}^*b_{k',\text{x}}\}\mathbb{E}\{a_{l,\text{x}}a_{l',\text{x}}^*\}+
\mathbb{E}\{b_{h,\text{y}}b_{k,\text{y}}^*b_{h',\text{y}}^*b_{k',\text{y}}\}\nonumber\\&\cdot\mathbb{E}\{a_{l,\text{x}}a_{l',\text{x}}^*\}+
\mathbb{E}\{b_{h,\text{x}}b_{k,\text{y}}^*b_{h',\text{x}}^*b_{k',\text{y}}\}\mathbb{E}\{a_{l,\text{y}}a_{l',\text{y}}^*\}
\Big).
\end{align}
We now focus on the calculation of the first term of \eqref{nli.variance.alt.reduced}, and we can compute the others in a similar way. To evaluate the fourth order moment given in the first term of \eqref{nli.variance.alt.reduced}, the following cases should be taken into account.
\begin{equation}\label{delta.fourth}
\begin{split}
&\mathbb{E}\{b_{h,\text{x}}b_{k,\text{x}}^*b_{h',\text{x}}^*b_{k',\text{x}}\}=\left\{
\begin{array}{rl}
\mathbb{E}\{|b_{\text{x}}|^4\}, & h=k=h'=k'\\
\mathbb{E}^2\{|b_{\text{x}}|^2\}, & h=k\neq h'=k'\\
\mathbb{E}^2\{|b_{\text{x}}|^2\}, & h=h'\neq k=k'.
\end{array}\right.
\end{split}
\end{equation}
Using \eqref{delta.fourth} and \eqref{X.coef.alt.freq}, 
we can write the contribution of the first term of \eqref{nli.variance.alt.reduced} to the NLI, as 
\begin{align}
    \label{second.term.cont}
  & \sigma_{\text{XPM}, \text{x},\text{1st}}^2\!=\!\frac{64}{81}\gamma^2\!\!\!\int\!\!\frac{\text{d}^3w}{(2\pi)^3}\frac{\text{d}^3w'}{(2\pi)^3} \rho_{\text{xp}}(w_1,w_2,w_3)\rho_{\text{xp}}^*(w_1',w_2',w_3')\nonumber\\&\Big(4\mathbb{E}\{|b_{\text{x}}|^4\}\mathbb{E}\{|a_{\text{x}}|^2\}\sum_{h}\text{e}^{i(w_1-w_2-w_1'+w_2')hT}\!\sum_{l}\!\text{e}^{i(w_3-w_3')lT}\nonumber\\&\!+\!
    4\mathbb{E}^2\{|b_{\text{x}}|^2\}\mathbb{E}\{|a_{\text{x}}|^2\}\!\Big[\!\sum_{h\neq h'}\!\text{e}^{i(w_1-w_2)hT-(w_1'-w_2')h'T}\sum_{l}\text{e}^{iw_3lT}\nonumber\\&\cdot\text{e}^{-w_3'lT}+\sum_{h\neq k}\text{e}^{i(w_1-w_1')hT-(w_2-w_2')kT}\sum_{l}\text{e}^{i(w_3-w_3')lT}\Big]\Big).
\end{align}
Considering \eqref{half.power}, \eqref{identity.delta} and \eqref{neq.double.sum}, we can express \eqref{second.term.cont} as
\begin{align}
    \label{second.term.cont.alt}
   &\sigma_{\text{XPM}, \text{x},\text{1st}}^2=\frac{8}{81}\gamma^2P_{\text{\tiny COI}}P_{\text{\tiny INT}}^2\int\frac{\text{d}^3w}{(2\pi)^3}\frac{\text{d}^3w'}{(2\pi)^3} \rho_{\text{xp}}(w_1,w_2,w_3)\nonumber\\&\cdot\rho_{\text{xp}}^*(w_1',w_2',w_3')\Big(\varphi_2\frac{16\pi^2}{T^2}\delta(w_1-w_2-w_1'+w_2')\nonumber\\&\cdot\delta(w_3-w_3')\!\!+\!
    4\Big[(\frac{8\pi^3}{T^3}\delta(w_1-w_2)\delta(w_1'-w_2')-\frac{4\pi^2}{T^2}\nonumber\\&\cdot\delta(w_1-w_2+w_1'-w_2'))\delta(w_3-w_3')+
(\frac{8\pi^3}{T^3}\delta(w_1-w_1')\nonumber\\&\cdot\delta(w_2-w_2')-\frac{4\pi^2}{T^2}\delta(w_1-w_2+w_1'-w_2'))\delta(w_3-w_3')\Big]\Big).
\end{align}
It should be noticed that the term $\delta(w_1-w_2)\delta(w_1'-w_2')$ is a bias term and should be ignored.
By excluding this term from \eqref{second.term.cont.alt}, we have
\begin{align}
    \label{second.term.final.variance}
    \sigma_{\text{XPM},\text{x},\text{1st}}^2=\frac{8}{81}\gamma^2P_{\text{\tiny COI}}P_{\text{\tiny INT}}^2\left[\left(\varphi_2-2\right)4X+4 Z\right],
\end{align}
where $X$ and $Z$ are given in Table~\ref{terms.identical}. 
Analogous expressions hold for other terms of \eqref{nli.variance.alt.reduced}. We can therefore express \eqref{nli.variance.alt.reduced} as \eqref{final.variance.total.main}.

\bibliographystyle{IEEEtran}
\bibliography{IEEEabrv,ref_beygi}

\begin{thebibliography}{10}
\providecommand{\url}[1]{#1}
\csname url@samestyle\endcsname
\providecommand{\newblock}{\relax}
\providecommand{\bibinfo}[2]{#2}
\providecommand{\BIBentrySTDinterwordspacing}{\spaceskip=0pt\relax}
\providecommand{\BIBentryALTinterwordstretchfactor}{4}
\providecommand{\BIBentryALTinterwordspacing}{\spaceskip=\fontdimen2\font plus
\BIBentryALTinterwordstretchfactor\fontdimen3\font minus
  \fontdimen4\font\relax}
\providecommand{\BIBforeignlanguage}[2]{{%
\expandafter\ifx\csname l@#1\endcsname\relax
\typeout{** WARNING: IEEEtran.bst: No hyphenation pattern has been}%
\typeout{** loaded for the language `#1'. Using the pattern for}%
\typeout{** the default language instead.}%
\else
\language=\csname l@#1\endcsname
\fi
#2}}
\providecommand{\BIBdecl}{\relax}
\BIBdecl

\bibitem{Essiambre_2010}
R.-J. Essiambre, G.~Kramer, P.~J. Winzer, G.~J. Foschini, and B.~Goebel,
  ``Capacity limits of optical fiber networks,'' \emph{J. Lightw. Technol.},
  vol.~28, no.~4, pp. 662--701, Feb. 2010.

\bibitem{betti1990exploiting}
S.~Betti, F.~Curti, G.~De~Marchis, and E.~Iannone, ``Exploiting fiber optics
  transmission capacity: 4-quadrature multilevel signalling,'' \emph{Electron.
  Lett.}, vol.~26, no.~14, pp. 992--993, July 1990.

\bibitem{betti1991novel}
------, ``A novel multilevel coherent optical system: 4-quadrature signaling,''
  \emph{J. Lightw. Technol.}, vol.~9, no.~4, pp. 514--523, Apr. 1991.

\bibitem{benedetto1992theory}
S.~Benedetto and P.~Poggiolini, ``Theory of polarization shift keying
  modulation,'' \emph{{IEEE} Trans. Commun.}, vol.~40, no.~4, pp. 708--721,
  Apr. 1992.

\bibitem{cusani1992efficient}
R.~Cusani, E.~Iannone, A.~M. Salonico, and M.~Todaro, ``An efficient multilevel
  coherent optical system: {M-4Q-QAM},'' \emph{J. Lightw. Technol.}, vol.~10,
  no.~6, pp. 777--786, June 1992.

\bibitem{Erik.optimized.modulation}
E.~{Agrell} and M.~{Karlsson}, ``Power-efficient modulation formats in coherent
  transmission systems,'' \emph{J. Lightw. Technol.}, vol.~27, no.~22, pp.
  5115--5126, Nov. 2009.

\bibitem{karlsson2009most}
M.~Karlsson and E.~Agrell, ``Which is the most power-efficient modulation
  format in optical links?'' \emph{Opt. Express}, vol.~17, no.~13, pp.
  10\,814--10\,819, June 2009.

\bibitem{Kojima2017}
K.~{Kojima}, T.~{Yoshida}, T.~{Koike-Akino}, D.~S. {Millar}, K.~{Parsons},
  M.~{Pajovic}, and V.~{Arlunno}, ``Nonlinearity-tolerant four-dimensional
  {2A8PSK} family for 5–7 bits/symbol spectral efficiency,'' \emph{J. Lightw.
  Technol.}, vol.~35, no.~8, pp. 1383--1391, Feb. 2017.

\bibitem{Reimer}
M.~{Reimer}, S.~O. {Gharan}, A.~D. {Shiner}, and M.~{O'Sullivan}, ``Optimized 4
  and 8 dimensional modulation formats for variable capacity in optical
  networks,'' in \emph{Proc. Optical Fiber Communication Conf.}, Anaheim, CA,
  USA, Mar. 2016.

\bibitem{Nakamura}
T.~{Nakamura}, E.~L.~T. {de Gabory}, H.~{Noguchi}, W.~{Maeda}, J.~{Abe}, and
  K.~{Fukuchi}, ``Long haul transmission of four-dimensional {64SP-12QAM}
  signal based on {16QAM} constellation for longer distance at same spectral
  efficiency as {PM-8QAM},'' in \emph{Proc. European Conf. Optical
  Communication}, Valencia, Spain, Sep. 2015.

\bibitem{Chen2019}
B.~{Chen}, C.~{Okonkwo}, H.~{Hafermann}, and A.~{Alvarado},
  ``Polarization-ring-switching for nonlinearity-tolerant geometrically shaped
  four-dimensional formats maximizing generalized mutual information,''
  \emph{J. Lightw. Technol.}, vol.~37, no.~14, pp. 3579--3591, July 2019.

\bibitem{Alvarado2015}
A.~{Alvarado} and E.~{Agrell}, ``Four-dimensional coded modulation with
  bit-wise decoders for future optical communications,'' \emph{J. Lightw.
  Technol.}, vol.~33, no.~10, pp. 1993--2003, May 2015.

\bibitem{Cai2020}
J.~{Cai}, M.~V. {Mazurczyk}, H.~G. {Batshon}, M.~{Paskov}, C.~R. {Davidson},
  Y.~{Hu}, O.~V. {Sinkin}, M.~A. {Bolshtyansky}, D.~G. {Foursa}, and A.~N.
  {Pilipetskii}, ``Performance comparison of probabilistically shaped {QAM}
  formats and hybrid shaped {APSK} formats with coded modulation,'' \emph{J.
  Lightw. Technol.}, vol.~38, no.~12, pp. 3280--3288, June 2020.

\bibitem{Frey2020}
F.~{Frey}, S.~{Stern}, J.~K. {Fischer}, and R.~F.~H. {Fischer}, ``Two-stage
  coded modulation for {Hurwitz} constellations in fiber-optical
  communications,'' \emph{J. Lightw. Technol.}, vol.~38, no.~12, pp.
  3135--3146, June 2020.

\bibitem{CarenaGaussian2012}
A.~Carena, V.~Curri, G.~Bosco, P.~Poggiolini, and F.~Forghieri, ``Modeling of
  the impact of nonlinear propagation effects in uncompensated optical coherent
  transmission links,'' \emph{J. Lightw. Technol.}, vol.~30, no.~10, pp.
  1524--1539, May 2012.

\bibitem{A.Mecozzi2012}
A.~Mecozzi and R.~J. Essiambre, ``Nonlinear {Shannon} limit in pseudolinear
  coherent systems,'' \emph{J. Lightw. Technol.}, vol.~30, no.~12, pp.
  2011--2024, June 2012.

\bibitem{Pontus_JLT_modeling_2013}
P.~Johannisson and M.~Karlsson, ``Perturbation analysis of nonlinear
  propagation in a strongly dispersive optical communication system,'' \emph{J.
  Lightw. Technol.}, vol.~31, no.~8, pp. 1273--1282, Apr. 2013.

\bibitem{dar2013properties}
R.~Dar, M.~Feder, A.~Mecozzi, and M.~Shtaif, ``Properties of nonlinear noise in
  long, dispersion-uncompensated fiber links,'' \emph{Opt. Express}, vol.~21,
  no.~22, pp. 25\,685--25\,699, Nov. 2013.

\bibitem{Curri2013}
V.~Curri, A.~Carena, P.~Poggiolini, G.~Bosco, and F.~Forghieri, ``{Extension
  and validation of the GN model for non-linear interference to uncompensated
  links using Raman amplification.}'' \emph{Opt. Express}, vol.~21, no.~3, pp.
  3308--17, Feb. 2013.

\bibitem{carena2014egn}
A.~Carena, G.~Bosco, V.~Curri, Y.~Jiang, P.~Poggiolini, and F.~Forghieri,
  ``{EGN} model of non-linear fiber propagation,'' \emph{Opt. Express},
  vol.~22, no.~13, pp. 16\,335--16\,362, June 2014.

\bibitem{Carena2012}
A.~Carena, V.~Curri, G.~Bosco, P.~Poggiolini, and F.~Forghieri, ``Modeling of
  the impact of nonlinear propagation effects in uncompensated optical coherent
  transmission links,'' \emph{J. Lightw. Technol.}, vol.~30, no.~10, pp.
  1524--1539, May 2012.

\bibitem{Poggiolini2012}
P.~Poggiolini, ``The {GN} model of non-linear propagation in uncompensated
  coherent optical systems,'' \emph{J. Lightw. Technol.}, vol.~30, no.~24, pp.
  3857--3879, Dec. 2012.

\bibitem{serena2013alternative}
P.~Serena and A.~Bononi, ``An alternative approach to the {Gaussian} noise
  model and its system implications,'' \emph{J. Lightw. Technol.}, vol.~31,
  no.~22, pp. 3489--3499, Nov. 2013.

\bibitem{Beygi_2012}
L.~Beygi, E.~Agrell, P.~Johannisson, M.~Karlsson, and H.~Wymeersch, ``A
  discrete-time model for uncompensated single-channel fiber-optical links,''
  \emph{{IEEE} Trans. Commun.}, vol.~60, no.~11, pp. 3440--3450, Nov. 2012.

\bibitem{Erik.Finite.memory.model}
E.~{Agrell}, A.~{Alvarado}, G.~{Durisi}, and M.~{Karlsson}, ``Capacity of a
  nonlinear optical channel with finite memory,'' \emph{J. Lightw. Technol.},
  vol.~32, no.~16, pp. 2862--2876, Aug. 2014.

\bibitem{poggiolini2016analytical}
P.~Poggiolini, A.~Nespola, Y.~Jiang, G.~Bosco, A.~Carena, L.~Bertignono, S.~M.
  Bilal, S.~Abrate, and F.~Forghieri, ``Analytical and experimental results on
  system maximum reach increase through symbol rate optimization,'' \emph{J.
  Lightw. Technol.}, vol.~34, no.~8, pp. 1872--1885, Apr. 2016.

\bibitem{secondini2013achievable}
M.~Secondini, E.~Forestieri, and G.~Prati, ``Achievable information rate in
  nonlinear {WDM} fiber-optic systems with arbitrary modulation formats and
  dispersion maps,'' \emph{J. Lightw. Technol.}, vol.~31, no.~23, pp.
  3839--3852, Dec. 2013.

\bibitem{Semrau.2019.MD.jlt}
D.~{Semrau}, E.~{Sillekens}, R.~I. {Killey}, and P.~{Bayvel}, ``A modulation
  format correction formula for the {Gaussian} noise model in the presence of
  inter-channel stimulated {Raman} scattering,'' \emph{J. Lightw. Technol.},
  vol.~37, no.~19, pp. 5122--5131, Oct. 2019.

\bibitem{rabbani2019general}
H.~Rabbani, G.~Liga, V.~Oliari, L.~Beygi, E.~Agrell, M.~Karlsson, and
  A.~Alvarado, ``A general analytical model of nonlinear fiber propagation in
  the presence of {Kerr} nonlinearity and stimulated {Raman} scattering,''
  {\it{arXiv}}, 2020. [Online]. Available:
  \url{http://arxiv.org/abs/1909.08714v2}.

\bibitem{Erik.Survey}
E.~{Agrell}, G.~{Durisi}, and P.~{Johannisson}, ``Information-theory-friendly
  models for fiber-optic channels: A primer,'' in \emph{IEEE Information Theory
  Workshop (ITW)}, Jerusalem, Israel, Apr.-May 2015.

\bibitem{golani2016modeling}
O.~Golani, R.~Dar, M.~Feder, A.~Mecozzi, and M.~Shtaif, ``Modeling the
  bit-error-rate performance of nonlinear fiber-optic systems,'' \emph{J.
  Lightw. Technol.}, vol.~34, no.~15, pp. 3482--3489, Aug. 2016.

\bibitem{liga2020extending}
G.~Liga, A.~Barreiro, H.~Rabbani, and A.~Alvarado, ``Extending fibre nonlinear
  interference power modelling to account for general dual-polarisation {4D}
  modulation formats,'' {\it{arXiv}}, Aug, 2020. [Online]. Available:
  \url{http://arxiv.org/abs/2008.11243}.

\bibitem{MecozziEssiambre2012}
A.~Mecozzi and R.-J. Essiambre, ``Nonlinear {Shannon} limit in pseudolinear
  coherent systems,'' \emph{J. Lightw. Technol.}, vol.~30, pp. 2011--2024, Jun.
  2012.

\bibitem{agrawal_linear_2006}
G.~P. Agrawal, \emph{Fiber-Optic Communication Systems}, 3rd~ed.\hskip 1em plus
  0.5em minus 0.4em\relax Wiley, 2002.

\bibitem{dar2015inter}
R.~Dar, M.~Feder, A.~Mecozzi, and M.~Shtaif, ``Inter-channel nonlinear
  interference noise in {WDM} systems: Modeling and mitigation,'' \emph{J.
  Lightw. Technol.}, vol.~33, no.~5, pp. 1044--1053, Mar. 2015.

\bibitem{Erik.database}
E.~{Agrell}, ``Database of sphere packings,'' 2014--2020. [Online]. Available:
  \url{http://codes.se/packings/}.

\bibitem{chen2020analysis}
B.~Chen, A.~Alvarado, S.~van~der Heide, M.~{van den Hout}, H.~Hafermann, and
  C.~Okonkwo, ``Analysis and experimental demonstration of orthant-symmetric
  four-dimensional 7 {bit/4D-sym} modulation for optical fiber communication,''
  {\it{arXiv}}, 2020. [Online]. Available:
  \url{http://arxiv.org/abs/2003.12712}.

\bibitem{Sjodin2013}
M.~{Sj\"odin}, E.~{Agrell}, and M.~{Karlsson}, ``Subset-optimized
  polarization-multiplexed {PSK} for fiber-optic communications,'' \emph{{IEEE}
  Commun. Lett.}, vol.~17, no.~5, pp. 838--840, May 2013.

\bibitem{tobias2015}
T.~A. {Eriksson}, S.~{Alreesh}, C.~{Schmidt-Langhorst}, F.~{Frey}, P.~W.
  {Berenguer}, C.~{Schubert}, J.~K. {Fischer}, P.~A. {Andrekson},
  M.~{Karlsson}, and E.~{Agrell}, ``Experimental investigation of a
  four-dimensional 256-ary lattice-based modulation format,'' in \emph{Proc.
  Optical Fiber Communication Conf.}, Los Angeles, CA, USA, Mar. 2015.

\bibitem{zetterberg1977codes}
L.~Zetterberg and H.~Br\"andstr\"om, ``Codes for combined phase and amplitude
  modulated signals in a four-dimensional space,'' \emph{{IEEE} Trans.
  Commun.}, vol.~25, no.~9, pp. 943--950, Sep. 1977.

\bibitem{carena2014accuracy}
A.~Carena, G.~Bosco, V.~Curri, Y.~Jiang, P.~Poggiolini, and F.~Forghieri, ``On
  the accuracy of the {GN}-model and on analytical correction terms to improve
  it,'' {\it{arXiv}}, 2014. [Online]. Available:
  \url{http://arxiv.org/abs/1401.6946}.

\bibitem{poggiolini2012detailed}
P.~Poggiolini, G.~Bosco, A.~Carena, V.~Curri, Y.~Jiang, and F.~Forghieri, ``A
  detailed analytical derivation of the {GN} model of non-linear interference
  in coherent optical transmission systems,'' {\it{arXiv}}, 2014. [Online].
  Available: \url{http://arxiv.org/abs/1209.0394}.

\end{thebibliography}

\end{document}